\documentclass[aps,prd,floats,nofootinbib]{revtex4}
\usepackage{graphicx}
\usepackage{amsmath}

\begin{document}

\title{The Spectral Shape and Photon Fraction as Signatures of the GZK Cut-Off}

\author{Andrew~M.~Taylor}
\affiliation{Max-Planck-Institut f\"ur Kernphysik, 
             Postfach 103980, D-69029 Heidelberg, GERMANY}
\author{Felix~A.~Aharonian}
\affiliation{Max-Planck-Institut f\"ur Kernphysik, 
             Postfach 103980, D-69029 Heidelberg, GERMANY}
\affiliation{Dublin Institute for Advanced Studies,
5 Merrion Square, Dublin 2, IRELAND}

\begin{abstract}
With the prospect of measuring the fraction of arriving secondary photons, produced through photo-pion 
energy loss interactions of ultra high energy cosmic ray (UHECR) protons with the microwave 
background during propagation, we investigate how information about the local UHECR source 
distribution can be inferred from the primary (proton)
to secondary (photon) ratio. As an aid to achieve this, we develop an analytic description for
both particle populations as a function of propagation time. 
Through a consideration of the shape of the GZK cut-off and the corresponding photon fraction
curve, we investigate the different results expected for both different maximum proton energies 
injected by the sources, as well as a change in the local
source distribution following a perturbative deformation away from a homogeneous description. 
At the end of the paper, consideration 
is made as to how these results are modified through extra-galactic magnetic field
effects on the proton's propagation. The paper aims to demonstrate how the shape of the cosmic ray
flux in the cut-off region, along with the photon fraction, are useful indicators of the cut-off
origin as well as the local UHECR source distribution.
\end{abstract}

\maketitle

\section{Introduction}

Ultra high energy cosmic rays with energies $\sim$10$^{19}$~eV arrive at Earth
with a frequency of less than $\sim$1~km$^{-2}$~yr$^{-1}$ (ie. with an energy flux of 30~eV~cm$^{-2}$~s$^{-1}$) 
in $\pi$ sr. The now complete Pierre Auger 
detector (Auger) \cite{Abraham:2004dt}, covering $\sim$3000~km$^{2}$ in area, is able to detect 
several hundred cosmic ray events per year at energies $\sim$10$^{19}$~eV. 
At energies around 10$^{20}$~eV, the arrival frequency reduces to below 0.01~km$^{-2}$~yr$^{-1}$, with
only a few events per year expected to be detectable by Auger. 
This demonstrates both the low luminosity of the arriving UHECR ``beam'' and the relatively good 
statistics achievable by Auger.
Thus, the arrival of these extremely large detectors allows more to be inferred than
simply the presence of the cut-off in the arriving UHECR spectrum.

Before their arrival, these UHECR must propagate across the astronomical distances between
their source and Earth. Though their composition remains unclear \cite{Unger:2007mc}, it seems reasonable to
assume that they contain a proton component. In this paper we will ignore a nuclei component
in our discussion of UHECR. A description of the propagation of heavier nuclear species up to iron 
can be found in ref.~\cite{Hooper:2006tn,Hooper:2008pm,Allard:2008gj}.
With the $p\gamma$ pair production and pion production cross-sections both rising quickly above the 
threshold values of 1~MeV and 145~MeV respectively (these photon energies being in the proton's rest-frame), 
once sufficient center-of-mass 
energy is available in the collisions, these processes become of relevance to UHECR protons during their
propagation to Earth, and they begin to lose energy through these particle production channels.

\begin{figure}[t!]
\begin{center}
\rotatebox{-90}{\resizebox{5.0cm}{!}{\includegraphics{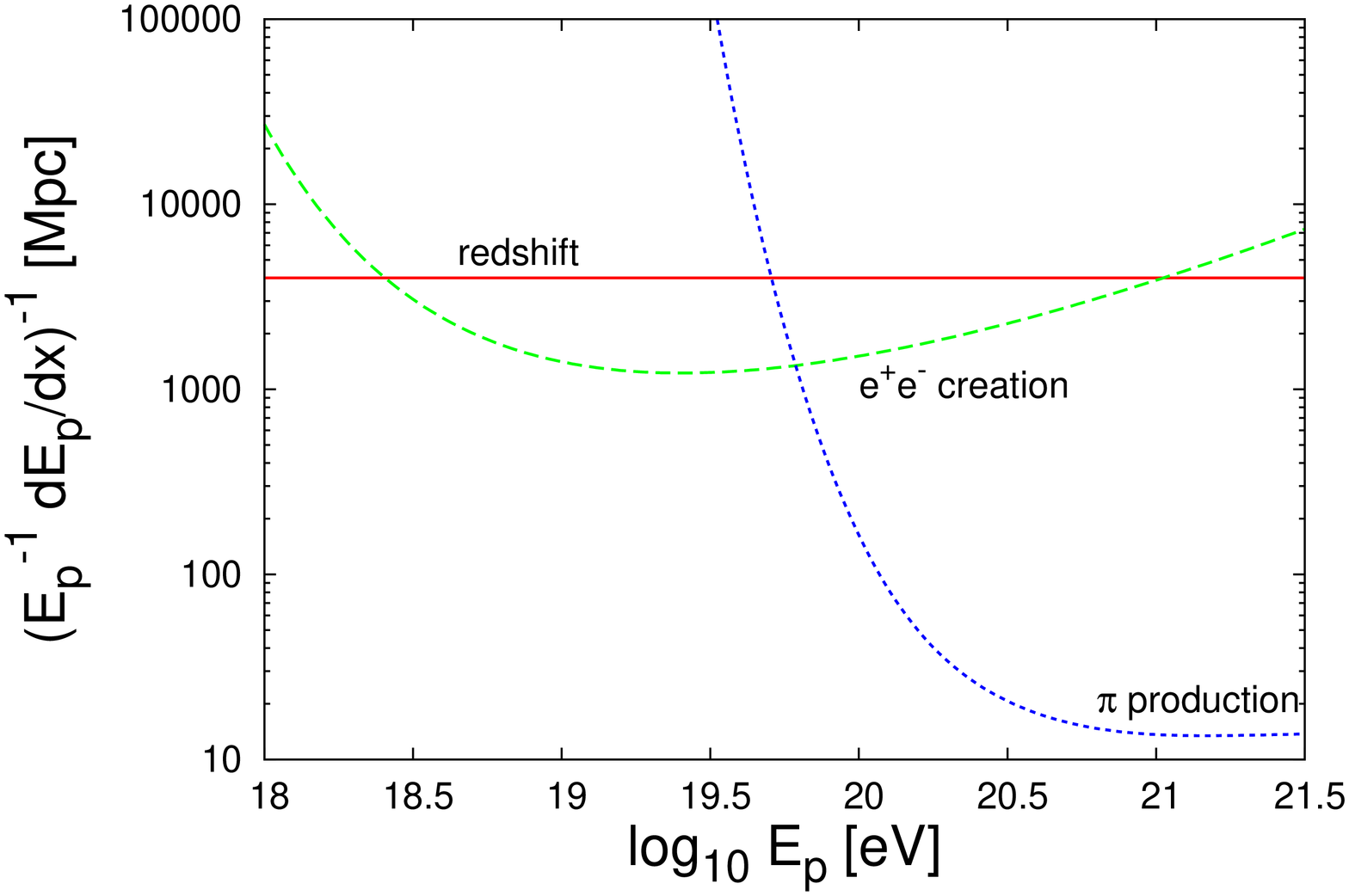}}}
\rotatebox{-90}{\resizebox{5.0cm}{!}{\includegraphics{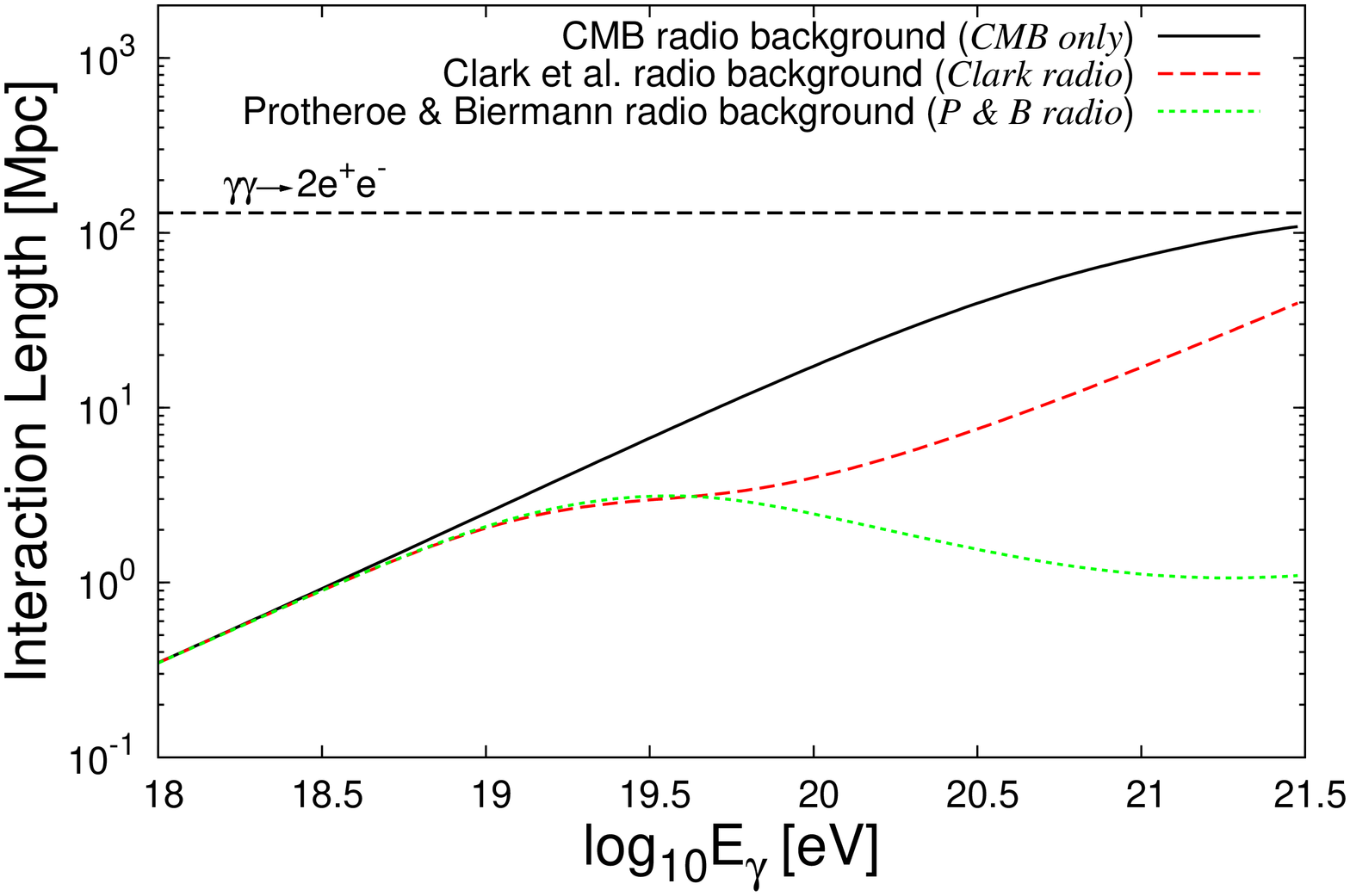}}}
\caption{{\bf Left}: The energy loss lengths lengths through redshift, pair creation, and pion production, for UHECR protons propagating through the cosmic microwave background radiation field. {\bf Right}: The interaction lengths for UHE photons propagating through the cosmic microwave and radio background radiation fields.}
\label{Energyloss_Times}
\end{center}
\end{figure}

At the high energy end of the UHECR spectrum, the required center-of-mass energy for both
pair ($p\gamma\rightarrow p e^{+}e^{-}$ \cite{Blumenthal:1970nn}) and pion production 
($p\gamma\rightarrow p \pi^{0}/n \pi^{+}$ \cite{Greisen:1966jv,Zatsepin:1966jv})
is found through collisions of the UHECR with both
2.7~K cosmic microwave background (CMB), whose energy density peaks with photons of energy $\sim$10$^{-3}$~eV, and to
a lesser extent with the infra-red background, whose energy density peaks with photons of energy $\sim$10$^{-2}$~eV. 
Provided the propagation time from their 
sources to Earth is greater than their energy loss time, the protons invariably undergo these energy loss 
interactions, generating a secondary flux of ultra high energy (UHE) electrons, photons, and neutrinos 
en-route, sharing out their energy flux among these various ``stable'' 
particle types via these particle production processes. In the left-panel of fig.~\ref{Energyloss_Times}, we 
show the calculated energy loss lengths for UHECR in the energy range under consideration, the 
determination of which is explained in ref.~\cite{Hooper:2006tn}.


Isotropy is observed in the arriving UHECR distribution detected by Auger for
energies below $\sim$10$^{19.7}$~eV, though above this energy there appears to be evidence 
(at the $\sim$3$\sigma$ level) for a departure from isotropy \cite{Abraham:2007si}.
Such a departure signal is naturally expected to occur in this energy range following the reduction 
of the UHECR energy loss lengths at these energies, approximately describable by \cite{Luca:inpress,Dermer:2007au}
\begin{eqnarray}
l_{\rm horiz.}=\frac{l_{0}e^{x}}{(1-e^{-x})}
\label{eq:horizon-rough}
\end{eqnarray}
where $l_{0}=5$~Mpc, $x=E_{p,0}/E_{p}$ and $E_{p,0}=10^{20.53}$~eV. This description gives a good approximation
to the proton photo-pion loss lengths, for interactions with the CMB only, in the range $1<x<10$.
With the UHECR above 10$^{19.6}$~eV arriving from increasingly more local sources, the overall deflection
angle of the arriving UHECR, from that at which they were emitted, is thus expected to reduce. 
Furthermore, for a given magnetic field strength, the UHECR
Larmor radius increases with proton energy, resulting in less deflection during propagation from a given source 
for higher energy particles.

Along with a correlation of the UHECR arrival directions with the local distribution of AGN, the shape 
of the GZK \cite{Greisen:1966jv,Zatsepin:1966jv} feature itself contains information about the source 
distribution \cite{Berezinsky:2002nc,Lu:2008rf}. 
As the cosmic rays produced by these sources are
attenuated via pion-production interactions with the background radiation field photons, a subsequent
flux of UHE photons, following the decay of the neutral pions, is produced.
However, unlike the flux of high energy (cosmogenic) neutrinos produced through the decay of the corresponding
charged pion component, previously discussed in ref~\cite{Hooper:2004jc,Ave:2004uj,Engel:2001hd}, UHE photons 
produced via the neutral pion decay are not free to propagate through the Universe over cosmological distances 
unimpeded since they also interact with the background photon field via 
$\gamma\gamma\rightarrow e^{+}e^{-}$ \cite{Gould:1967} on $\sim$Mpc size scales 
(see right-hand panel of fig.\ref{Energyloss_Times}), 
thus their flux is attenuated on scales similar to that on which they are 
produced through high center-of-mass $\gamma\gamma$ collisions.
These UHE electrons, produced through $\gamma\gamma$ pair production, may then go on to either inverse 
Compton cool \cite{Blumenthal:1970} off the background photon fields through high center-of-mass
$e\gamma$ collisions, leading to a repeating cycle of pair production and inverse
Compton scattering and the development of an electromagnetic ($e/\gamma$) cascade 
\cite{Protheroe:1986, Aharonian:1990b,Protheroe:1992dx}, 
or to synchrotron cool in the extra-galactic magnetic field. The dominance of one of these processes depending 
on both the level of the cosmic radio background and the magnitude of the extra-galactic magnetic field 
\cite{Aharonian:1992qf,Gelmini:2005wu}.

With no prior knowledge about the UHECR source population, a vanilla sky homogeneous and isotropic
distribution is naturally assumed. We use such an assumption as our starting point in this paper, departing
from this assumption in a perturbative way so as to explore the effects introduced by such departures into 
the arriving proton and photon fluxes.
With $>10^{18}$~eV protons energy loss scales being 10-1000~Mpc and $>10^{18}$~eV photons energy loss 
scales being 1-100~Mpc, and source distance scales expected $\sim$50~Mpc, the UHECR proton and photon 
fluxes collectively are perfectly suited to use as a probe of the local source distribution.

In order to quantify the effects of different source distributions in this paper, we separate
out the fluxes produced from source regions with shells of radii 0-3~Mpc, 3-9~Mpc, 9-27~Mpc,
27-81~Mpc, and 81-243~Mpc surrounding the Earth. In this way, the results obtained may be used
to encapsulate the effects introduced 
by a local over-density or under-density of UHECR sources.
As the ultimate fate of the energy flux injected into the UHE photon population by our UHECR 
population is dependent upon both the strength of the background radiation field at radio energies, 
10$^{-8}$-10$^{-6}$~eV, and the extra-galactic magnetic field strength, we account for the different 
eventualities the energy propagation through the system may follow through a consideration 
of extreme cases.

Previous investigations into the photon flux produced through GZK interactions
\cite{Aharonian:1990b,Aharonian:1992qf,Gelmini:2005wu}, have demonstrated that significant photon 
fractions can be expected for
particular assumptions about the UHECR sources and extra-galactic environment.
We demonstrate here that the photon fraction naturally provides a complimentary measurement to that of the 
spectral cut-off feature, allowing differentiation between the ``tired source'' and GZK cut-off scenarios
as well as carrying information about the distribution of the local UHECR source population. We aid
our discussion with the development of an analytic description of the proton and photon fluxes, which
provides a clear understanding of our results.
With Auger measurements of the arriving UHECR flux already probing the photon fraction down to values
$\sim$10$^{-2}$, and the possibility to probe even smaller fractions \cite{Aglietta:2007yx}, the crucial 
complimentary information this will provide to the spectral UHECR measurements holds great promise.

In section~\ref{source_shells} we consider the contribution to both the arriving UHECR
flux and arriving UHE photon flux from different source shells. 
An analytic description for both these fluxes is derived.
We go on to highlight the dependence of the photon flux
on the cosmic radio background and extra-galactic magnetic field strengths assumed. 
In section~\ref{Photon_Fraction} we obtain the photon fraction for the case of a homogeneous
distribution of sources. We then demonstrate the role that the cut-off energy, $E_{\rm max}$,
plays in this photon fraction result. Following this we investigate 
the effect different local source distributions would have on the spectral-shape of the cut-off and the photon 
fraction of the arriving UHECR flux. The section ends with the consideration of
the extreme case of the development with distance of the photon fraction from a single source. 
In section~\ref{diffusive_effects} we discuss how diffusion 
of the UHECR would affect the results obtained. We discuss a summary of our results in 
section~\ref{Discussion}, making our conclusions in section~\ref{Conclusion}.

\section{Proton and Photon Fluxes from a Uniform Distribution of Sources}
\label{source_shells}

\begin{figure}[t]
\begin{center}
\rotatebox{-90}{\resizebox{5.0cm}{!}{\includegraphics{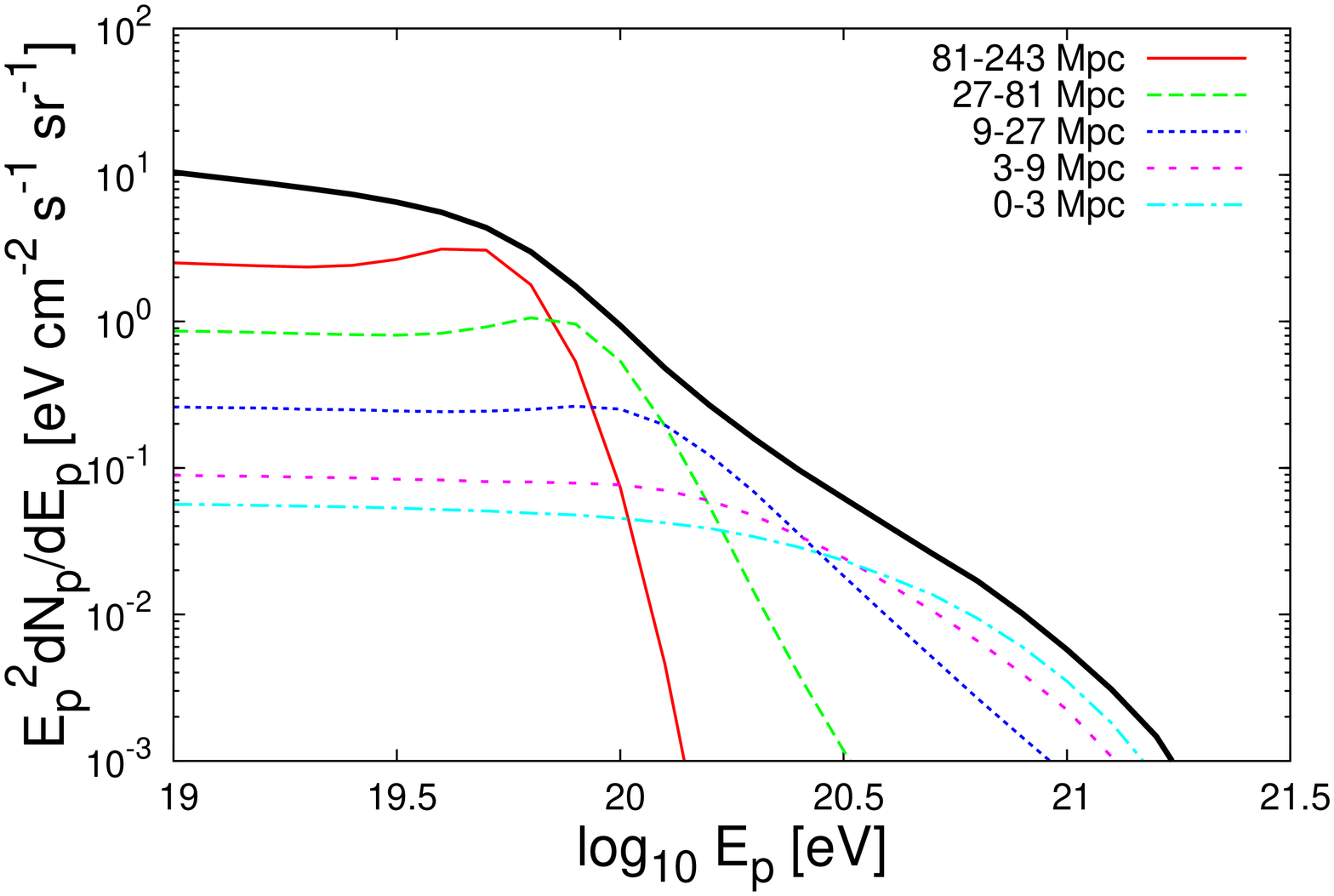}}}
\rotatebox{-90}{\resizebox{5.0cm}{!}{\includegraphics{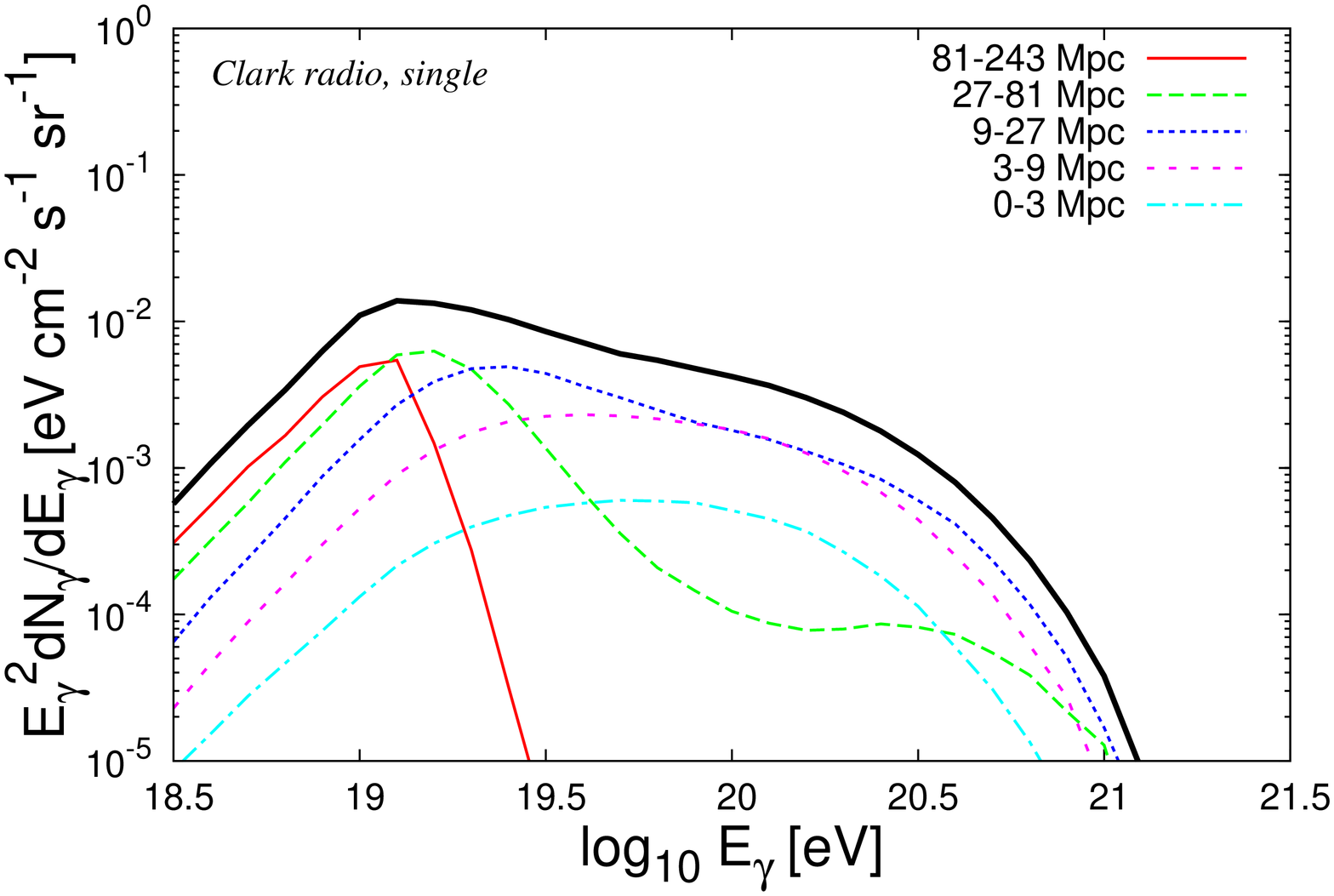}}}
\caption{{\bf Left}: A breakdown of the arriving UHECR proton flux from different injection shell radii. {\bf Right}: A breakdown of the arriving UHE photon flux from different injection shell radii. For this flux, the label ``{\it Clark radio, single}'' refers to the assumptions about the radio background and magnetic field strengths, for which we have assumed the Clark et al. \cite{Parker:1970} radio background and extra-galactic magnetic fields stronger than 0.1~nG. The injected proton spectrum assumed for these results had a cut-off energy of $E_{\rm max}$=10$^{20.5}$~eV and an injection spectral index of $\alpha$=2.}
\label{Ensemble_Results_Breakdown}
\end{center}
\end{figure}

In this section we assume that the UHECR sources have a homogeneous density distribution locally, with the number 
of sources scaling simply with the co-moving volume element under consideration. Along with this, we assume 
each source is producing UHECR with a 
power law energy spectrum and an exponential cut-off. These source spatial and energy distributions are given by
\begin{align}
\frac{dN}{dV_{c}}\propto (1+z)^{3}&&\frac{dN}{dE}\propto E^{-\alpha}e^{-E/E_{\rm max}}
\end{align}
respectively, with $\alpha$=2 and $E_{\rm max}=10^{20.5}$~eV, unless otherwise stated. Following these assumptions, 
the arriving proton and photon fluxes are shown in fig.~\ref{Ensemble_Results_Breakdown}.

The proton fluxes from the different source shells, shown in the left-panel of 
fig.~\ref{Ensemble_Results_Breakdown}, demonstrate simply how each shell has an energy range at 
which it contributes the dominant flux component of the arriving UHECR. Accompanying the arriving
proton flux, we also show the arriving photon flux in the right-panel of fig.~\ref{Ensemble_Results_Breakdown}, 
produced en-route after the protons have left
the source region, typically after a distance $l_{\rm horiz.}$ (see eqn~(\ref{eq:horizon-rough})). 
For the photon flux arriving from
the more distant shells ($>$27~Mpc), a clear dip feature can be seen to exist at energies of $\sim$10$^{20}$~eV.
This feature originates from the sudden increase in the pion production rate $\sim$10$^{19.6}$~eV, as seen in 
dotted energy-loss length curve in the left-panel of fig.~\ref{Energyloss_Times}, occurring when the protons 
have sufficient Lorentz factors that the populous CMB photons have sufficient energy to excite their delta 
resonances. Since, when this occurs, the average distance at which the pions are produced from the emitting 
shell drops as $e^{E_{p,0}/E_{p}}$, the source region disappears behind the pair production horizon leading to 
the sudden falling part seen before the dip feature. At higher energies the pair production horizon increases 
leading to the slower rising part after the dip feature.

\subsection{Analytic Description for the UHE Proton and Photon Fluxes}
\label{Analytic_Section}

The shape of the GZK feature, the shell contributions to which are shown in the left-hand panel of 
fig.~\ref{Ensemble_Results_Breakdown}, carries valuable information about the local source
distribution. A simple analytic description of this feature, found in 
ref.~\cite{Aharonian:1990,Karakula:1994nv}, allows the different cut-offs in the fluxes from the various 
shells to be interpreted easily. The derivation of this analytic result is explained in 
ref.~\cite{Hooper:2008pm}

The contribution function, for protons contributing to the arriving flux with energy $E_{p}$, 
written as a function of the total propagation distance $L$ from the source, for proton's 
having already undergone $n$ pion production interactions, but
which have yet to undergo their $(n+1)^{\rm th}$ pion production process, is given by,
\begin{eqnarray}
\frac{N_{n} (E_{p},L)}{N_{0} (E,0)} = \sum_{m=0}^{n} l_{0} l_m^{n-1} 
\exp\left(-{\frac{L}{l_m}}\right)\prod_{p=0(\neq m)}^{n} \frac{1}{l_m -l_p}
\label{distribution} 
\end{eqnarray}
where $l_{m}=l(E_{p}/(1-K_{p})^{m})$ is the pion production interaction length for a proton of energy 
$E_{p}/(1-K_{p})^{m}$. $K_{p}$, the pion production inelasticity, may be approximated as
$K_{p}=1/2(1-(m_{p}^{2}c^{4}-m_{\pi}^{2}c^{4})/s)$ \cite{Stecker:1968}, for the energy range
of interest, where $s$ is the invariant squared center-of-mass energy of the collision, 
$m_{p}$ is the proton rest-mass, and $m_{\pi}$ is the pion rest-mass. 
A simple analytic expression describing $l(E_{p})$ for proton interactions with
the CMB alone is given in eqn~(\ref{eq:horizon-rough}) with $l_{0}=1$~Mpc, $x=E_{p,0}/E_{p}$ and 
$E_{p,0}=10^{20.53}$~eV, which provides a good description up to proton energies $\sim$10$^{20.5}$~eV.

\begin{figure}[t]
\begin{center}
\rotatebox{-90}{\resizebox{5.0cm}{!}{\includegraphics{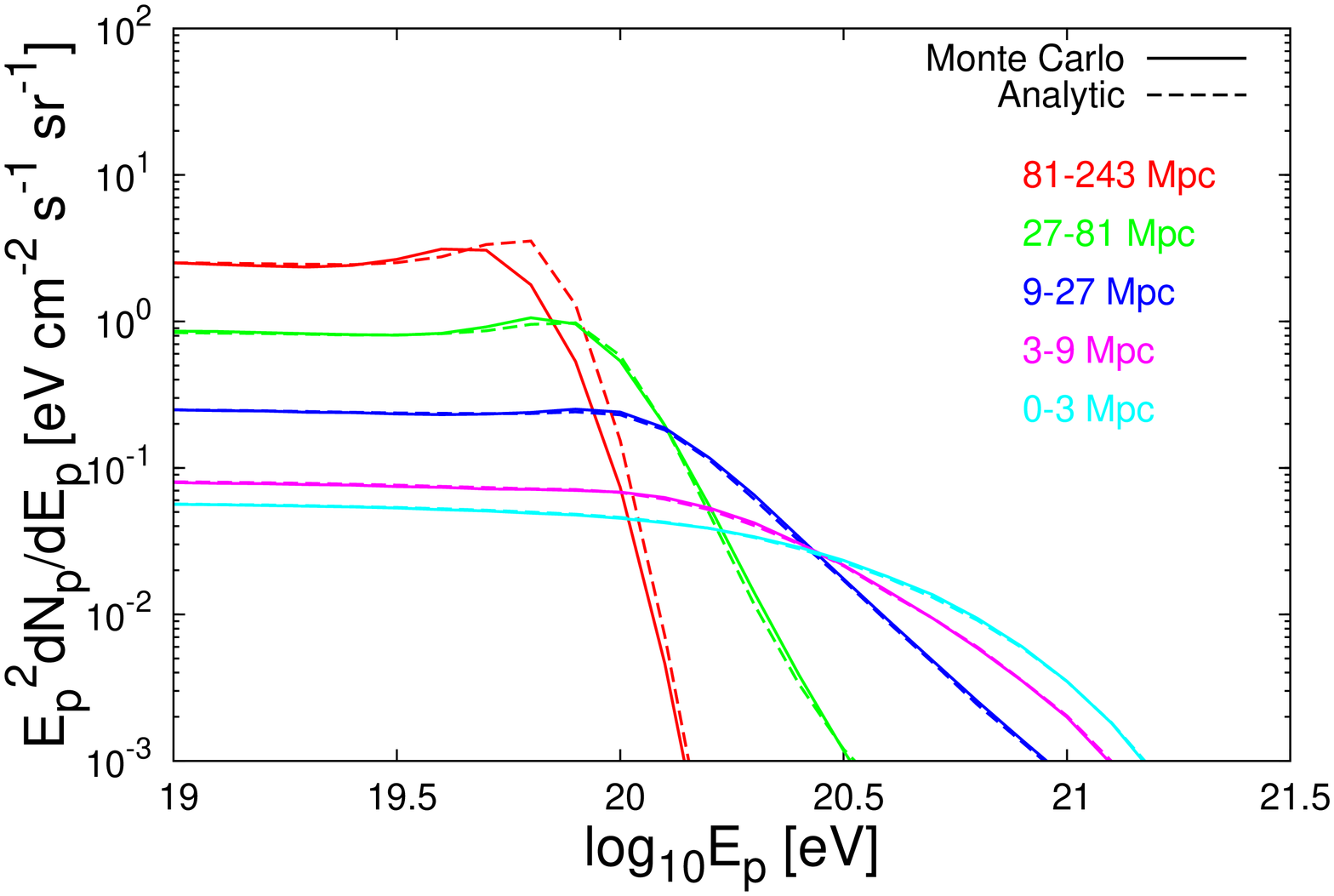}}}
\rotatebox{-90}{\resizebox{5.0cm}{!}{\includegraphics{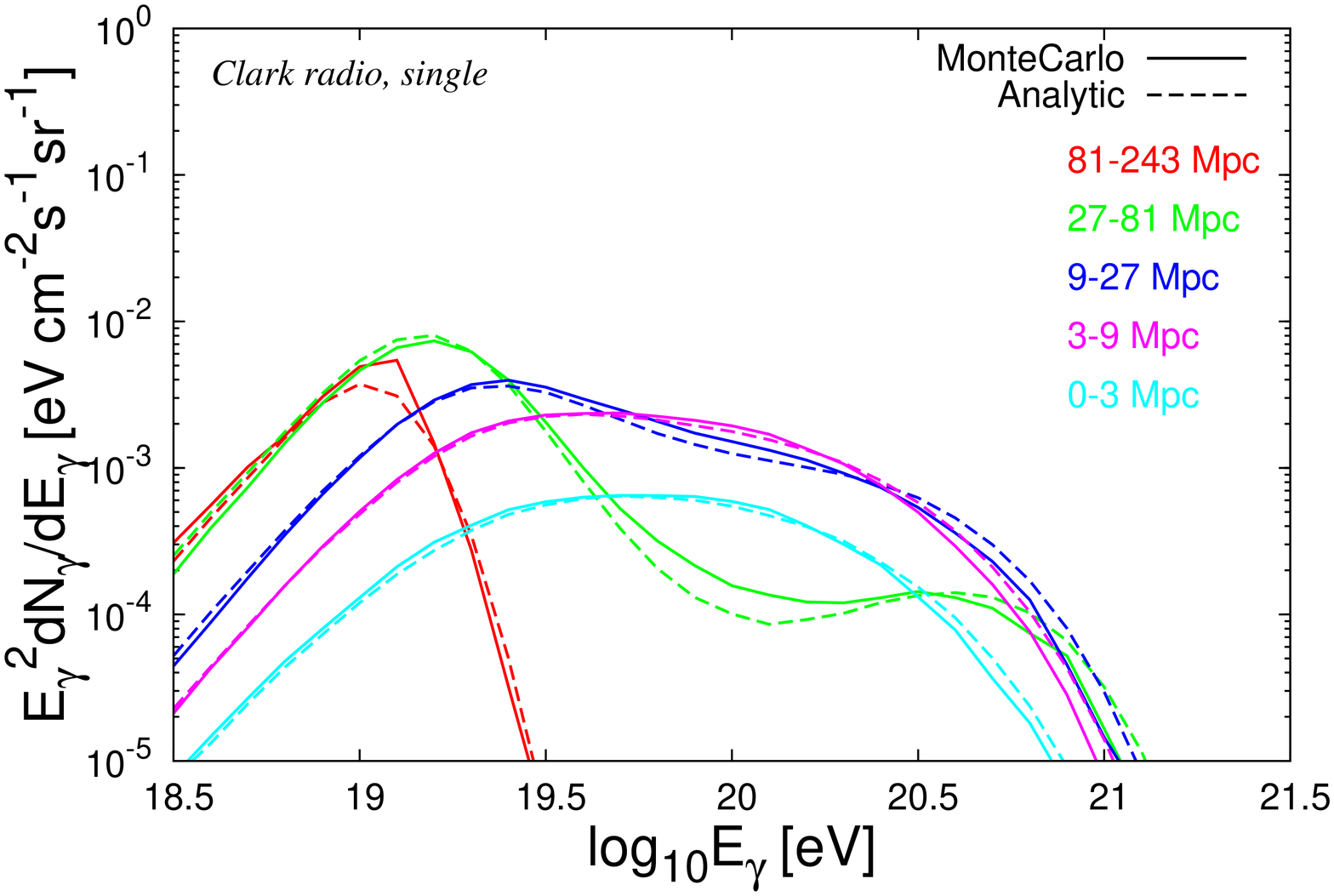}}}
\caption{{\bf Left}: A comparison of the analytic description for the GZK feature of the proton flux with the complete Monte Carlo description. {\bf Right}: A comparison of the analytic description given above for the photon flux produced during UHECR proton propagation with the complete Monte Carlo description. For this flux, the label ``{\it Clark radio, single}'' refers to the assumptions about the radio background and magnetic field strengths, for which we have assumed the Clark et al. \cite{Parker:1970} radio background and extra-galactic magnetic fields stronger than 0.1~nG. To obtain both sets of analytic results the formalism in section~\ref{Analytic_Section} was used. The injected proton spectrum assumed for these results had a cut-off energy of $E_{\rm max}$=10$^{20.5}$~eV and an injection spectral index of $\alpha$=2.}
\label{Analytic}
\end{center}
\end{figure}

Combining expressions (\ref{eq:horizon-rough}) and (\ref{distribution}), summing over $n_{\rm max}$
different pion loss contribution functions for the flux from the source, the total proton flux from 
a single source, at distance $L$ is,
\begin{eqnarray}
\frac{dN_{\rm total} (E_{p},L)}{dL}=\sum_{n=0}^{n_{\rm max}}\frac{N_{n} (E_{p},L)}{N_{0} (E,0)}.
\end{eqnarray} 
Integrating $L$ over the relevant source distribution distance range, the total proton flux from
the UHECR source production region is obtained.
This gives a reasonable analytic description of the GZK-feature, as demonstrated in 
the left-panel of fig.~\ref{Analytic}. The discrepancy between the analytic and Monte Carlo 
description, seen in the plot, for protons from the most distant shell
(81-243~Mpc) being due to the neglection of pair losses whose role becomes more important when $>$100~Mpc
propagation distances are probed. This can be understood from the ``$e^{+}e^{-}$ creation'' curve in 
the left-panel of fig.~\ref{Energyloss_Times} which demonstrates that loss lengths of a few 1000~Mpc, 
through proton pair losses in the CMB, are expected.
Since the injection spectral index, $\alpha$, of the UHECR produced by the sources only comes into the 
calculation of the arriving flux through the weighting factor in the sum over the different pion loss 
contributions (ie. the different $m$ in eqn~(\ref{distribution})), the spectral index is found to play a 
very minor role in the spectral slope of the GZK feature.
Hence, the effect of a steeper spectral index on the arriving proton spectra is simply to steepen
the arriving flux at energies below the GZK feature. Thus, with a smaller energy flux of protons injected
into the UHE-photon population in the GZK energy range, a reduced energy flux is injected into the UHE-photon 
population at the energies around the beginning of the cut-off (10$^{19.6}$~eV), with the injected energy flux 
reduced further still at even higher photon energies.

A calculation of the arriving photon spectrum produced via neutral photo-pion production 
interactions of the UHECR protons is a little more involved, requiring the consideration of
their production as well as their attenuation through subsequent pair production interactions.
In analogy with the production of secondary protons for photo-nuclear disintegration given in 
ref.~\cite{Hooper:2008pm}, the spectrum of neutral pions produced through photo-pion production after
having propagated distance $L'$ from the source is given by,
\begin{eqnarray}
N_{\pi}(L',K_{p}E_{p})=\int_{0}^{L'}dL \sum_{n=0}^{n_{\rm max}}\frac{N_{n} (E_{p},L)}{l_{n}}.\label{pion_contribution}
\end{eqnarray}
Hence, the spatial distribution of neutral pions produced through the propagation of an ensemble of
given energy protons, over the entire range in $L$ of source distances under consideration, is obtained. 
A comparison of the arriving photon flux (following the rapid decay of the neutral pions to photons) 
from the Monte Carlo and analytic descriptions is shown in the right-hand panel of fig.~\ref{Analytic}.

\subsection{The Impact of Different Cosmic Radio Background and Extra-galactic Magnetic Field Strengths}
\label{magentic_radio}

\begin{figure}[t]
\begin{center}
\rotatebox{-90}{\resizebox{5.0cm}{!}{\includegraphics{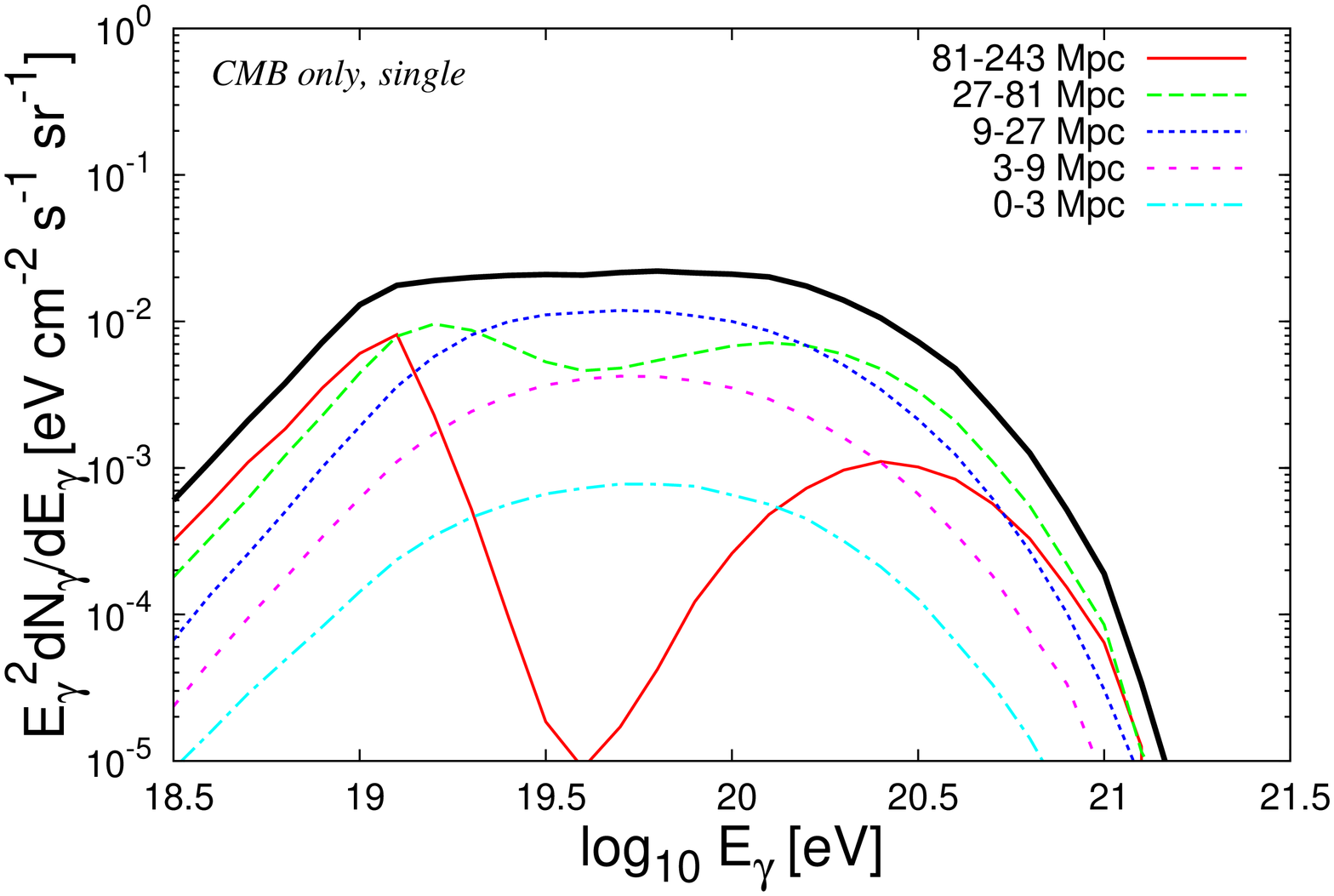}}}
\rotatebox{-90}{\resizebox{5.0cm}{!}{\includegraphics{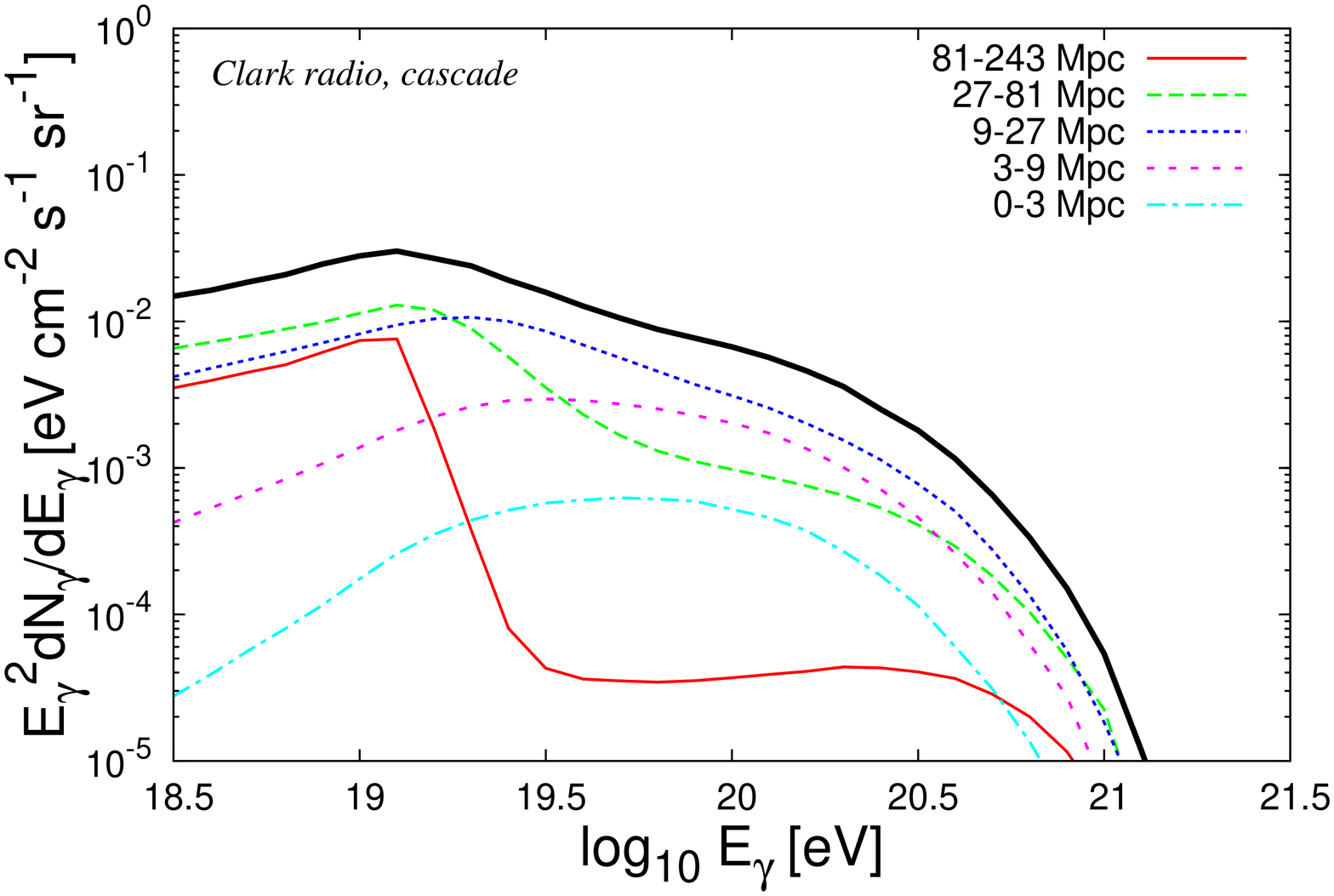}}}
\caption{{\bf Left:} A breakdown of the arriving UHECR photon flux from different injection shell radii for the case of a low radio background and relatively strong magnetic field, ``{\it CMB only, single}''. {\bf Right:} A breakdown of the arriving UHECR photon flux from different injection shell radii of for the case of a relatively strong radio background and weak magnetic field, ``{\it Clark radio, cascade})''. The injected proton spectrum assumed for these results had a cut-off energy of $E_{\rm max}$=10$^{20.5}$~eV and an injection spectral index of $\alpha$=2.}
\label{Ensemble_Results_Breakdown2}
\end{center}
\end{figure}

In the calculations presented in the right-panel of fig.~\ref{Ensemble_Results_Breakdown}, 
the propagation of the energy in secondaries soley in the form of photons through the
system was implicitly assumed, with the energy of the pair creation electrons 
disappearing from the system through the synchrotron channel.
However, with the present degree of uncertainty in the strengths of the 
cosmic radio background and extra-galactic magnetic fields, the favoured route that this energy takes
also remains uncertain.
Following the photon fluxes obtained in the previous section, we here demonstrate the variation introduced 
by either a weaker radio background or weaker extra-galactic magnetic field strength. 

We consider the extreme cases of: 
a low radio background (dominated by the low energy CMB) and single pair production after 
which the energy disappears from the system through the synchrotron channel (considering throughout only 
photon energies above 10$^{18}$~eV); 
a large radio background \cite{Parker:1970} and single pair production with electrons cooling via the
synchrotron channel; a large radio background and full $e/\gamma$ cascading. For the second of these,
the electron inverse Compton interaction lengths and both electron inverse Compton and photon pair production 
inelasticities were calculated \cite{Aharonian:1981,Aharonian:1983} and a Monte Carlo description of
the cascade was used.
We refer to this radio background measurement result as large due to possible local foreground contamination
from synchrotron emission in the Milky Way's halo.
Following radio background measurements, nearly 40 years ago, attempts have been made to model total 
radio background emitted by sources \cite{Protheroe:1996si}.
However, for the case of a weaker radio background, we consider the extreme case for which the low 
energy (radio) CMB component is the dominant contribution. Though this scenario is certainly extreme, we use it 
here simply to demonstrate the effects from a range of parameter space. We refer to this scenario 
as ``{\it CMB only, single}'', presenting the results in the left-hand panel of 
fig.~\ref{Ensemble_Results_Breakdown2}. As in the previous section for which the 
``{\it Clark radio, single}'' scenario was used, a dip feature is seen in the arriving photon flux from 
the more distance shells ($>$27~Mpc). The origin of the feature is the same as that described previously 
(see the beginning of section \ref{source_shells}).
For the case of a weaker extra-galactic magnetic field, we consider the case of sufficiently weak magnetic 
fields such that the $e/\gamma$ cascade's development is unhindered.
In this scenario, synchrotron cooling times of UHE electrons are slower than the inverse 
Compton cooling times on the background radiation fields. With synchrotron cooling times going as, 
$c\tau_{\rm sync.}= 400~{\rm Mpc}\left(\frac{10^{-11}{\rm G}}{B}\right)^{2}\left(\frac{10^{20}~{\rm eV}}{E_{e}}\right)$,
magnetic fields of $<$10$^{-11}$~G are sufficiently weak that these synchrotron cooling times become much larger than 
the 0.3~Mpc ($\sim$1~Myr) interactions lengths (for which the inelasticity per interaction is near unity).
Once again, we consider such extreme scenarios in order to capture the effect from a range of 
parameter space. We refer to this scenario as ``{\it Clark radio, cascade}'', presenting the results in the 
right-hand panel of fig.~\ref{Ensemble_Results_Breakdown2}. Furthermore, for this scenario, the development of 
the $e/\gamma$ cascades leads to the filling in, and partial erasing, of the dip feature seen previously in the 
non-cascading result shown in the right-hand panel of fig.~\ref{Ensemble_Results_Breakdown} in 
section~\ref{source_shells}.

\section{The Photon Fraction}
\label{Photon_Fraction}

\begin{figure}[t]
\begin{center}
\rotatebox{-90}{\resizebox{5.0cm}{!}{\includegraphics{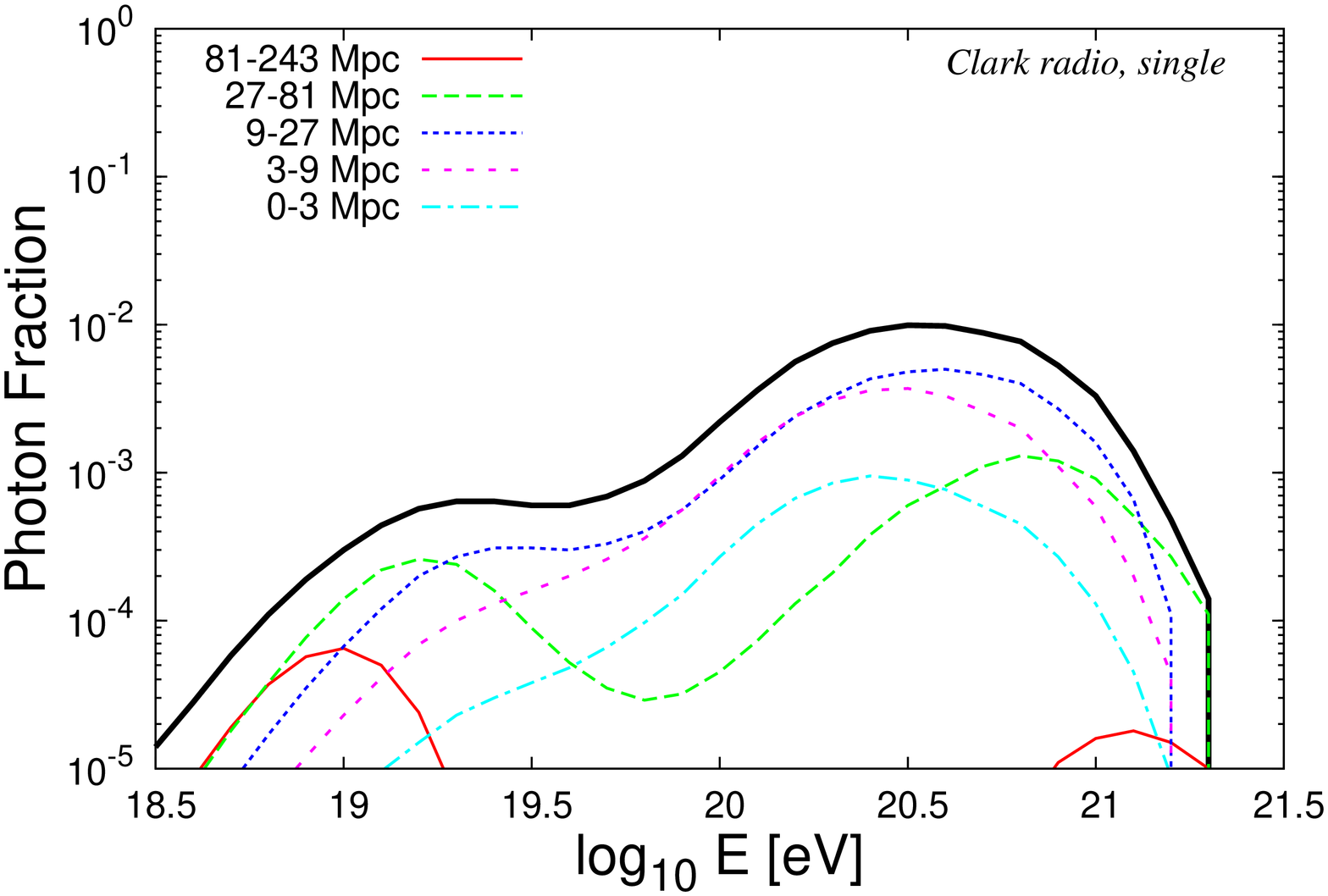}}}
\rotatebox{-90}{\resizebox{5.0cm}{!}{\includegraphics{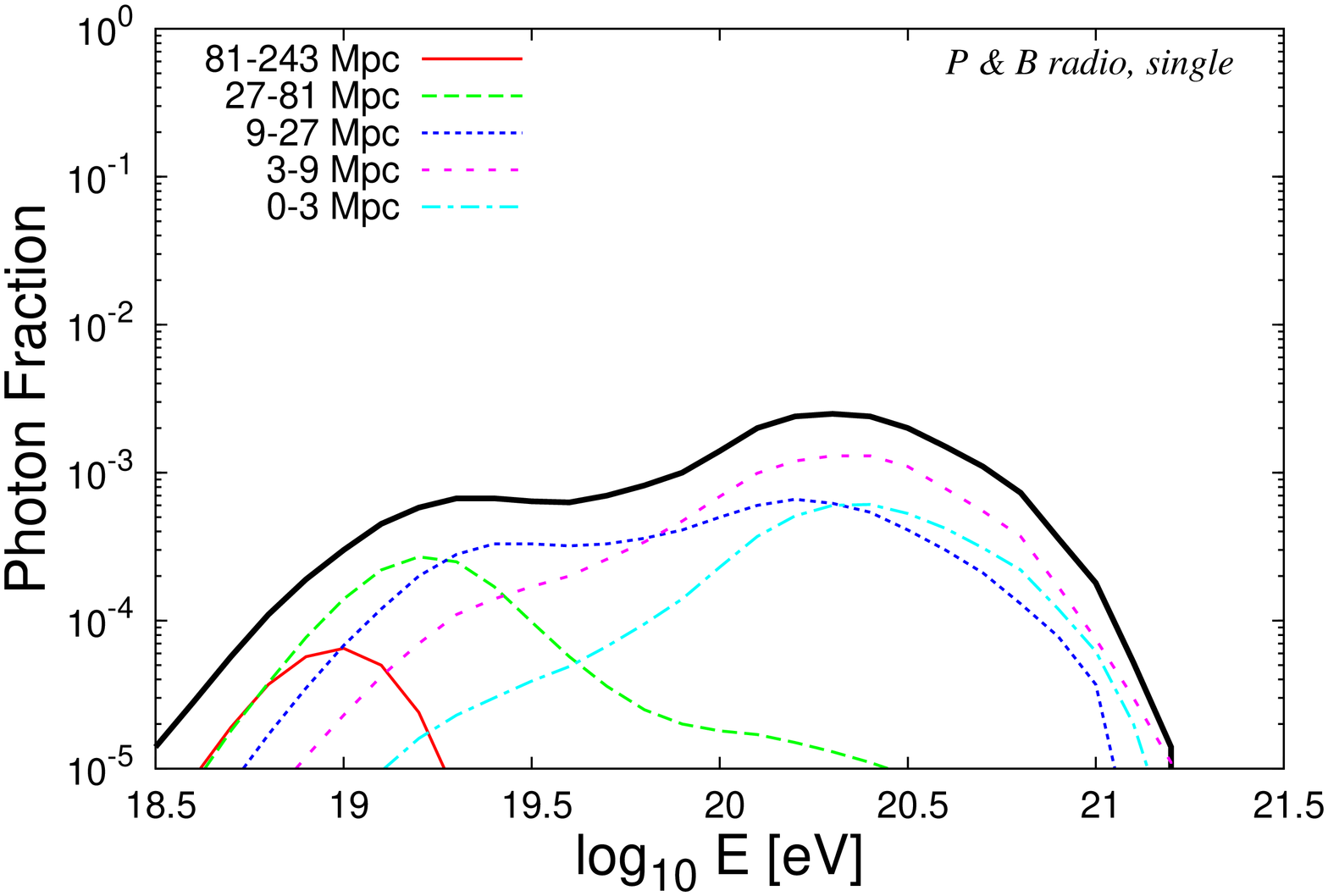}}}

\caption{{\bf Left:} A breakdown of the arriving UHE photon fraction. The label ``{\it Clark radio, single}'' refers to the assumptions about the radio background and magnetic field strengths, for which we have assumed the Clark et al. \cite{Parker:1970} radio background and extra-galactic magnetic fields stronger than 0.1~nG. {\bf Right:} A breakdown of the arriving UHE photon fraction.  The label ``{\it Protheroe \& Biermann, single}'' indicates  the the Protheroe \& Biermann radio background \cite{Protheroe:1996si} and extra-galactic magnetic fields stronger than 0.1~nG.For these plots a homogeneous distribution of sources were assumed, each source outputting a proton spectrum with a cut-off energy of $E_{\rm max}$=10$^{20.5}$~eV and an injection spectral index of $\alpha$=2.}
\label{Photon_Fractions}
\end{center}
\end{figure}

Since the quantity usually obtained from searches for UHE photons in UHECR data
is the photon fraction \cite{Aglietta:2007yx}, we present the photon fraction calculated for the 
``{\it Clark radio, single}'' scenario in fig.~\ref{Photon_Fractions}, obtained by dividing the photon and 
proton fluxes shown in the two panels of fig.~\ref{Ensemble_Results_Breakdown}. 
In this plot we also present a breakdown of the contributing photon
fluxes to the photon fraction from the different shells. 
From this breakdown it can be seen that the arriving photon
fraction is predominantly contributed from sources in relatively local shells, with distances
tens of Mpc away.

The dip feature seen in the photon fraction in fig.~\ref{Photon_Fractions} originates from a convolution of
the GZK feature in the proton flux itself along with the dip feature present in the arriving
photon flux explained previously in section \ref{Ensemble_Results_Breakdown}. For the homogeneous
source distribution case, photon fractions between 10$^{-3}$ and 10$^{-2}$ can be seen to be
expected, with the photon fraction peaking at 10$^{-2}$ close to the maximum energy the cosmic rays are
injected up to before the cutoff kicks in at energy $E_{\rm max}$.
These results are found to be in approximate agreement with previous calculations of the photon
fraction \cite{Gelmini:2005wu,Kalashev:2007sn}.
However, this photon fraction result has been calculated for a single maximum energy, $E_{\rm max}$, 
of the protons produced by the UHECR sources. In the following section we address the influence 
other values of $E_{\rm max}$ have on these results.


\subsection{The Effect of Different Cut-Off Energies}
\label{cut-off_effect}

\begin{figure}[t]
\begin{center}
\rotatebox{-90}{\resizebox{5.0cm}{!}{\includegraphics{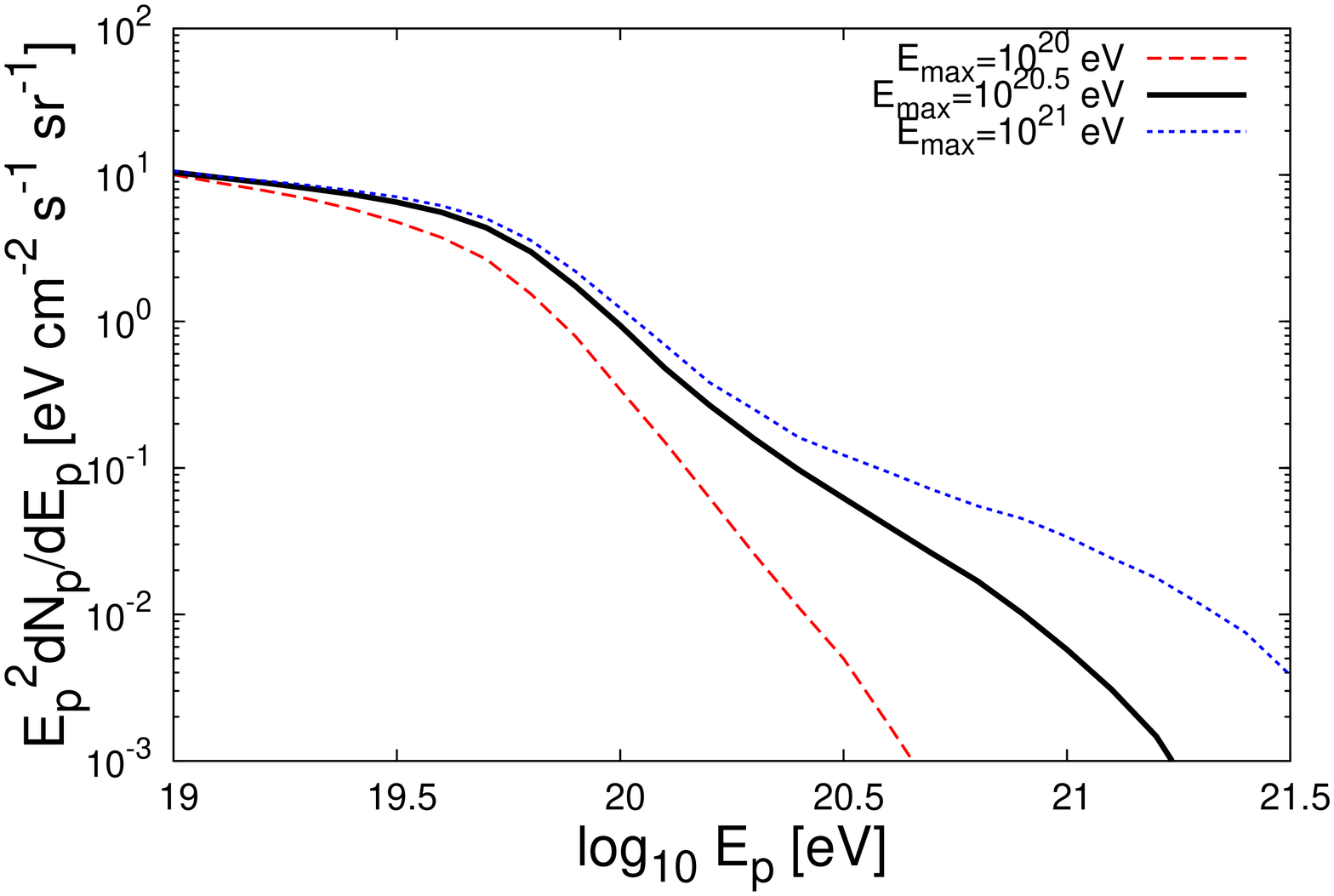}}}
\rotatebox{-90}{\resizebox{5.0cm}{!}{\includegraphics{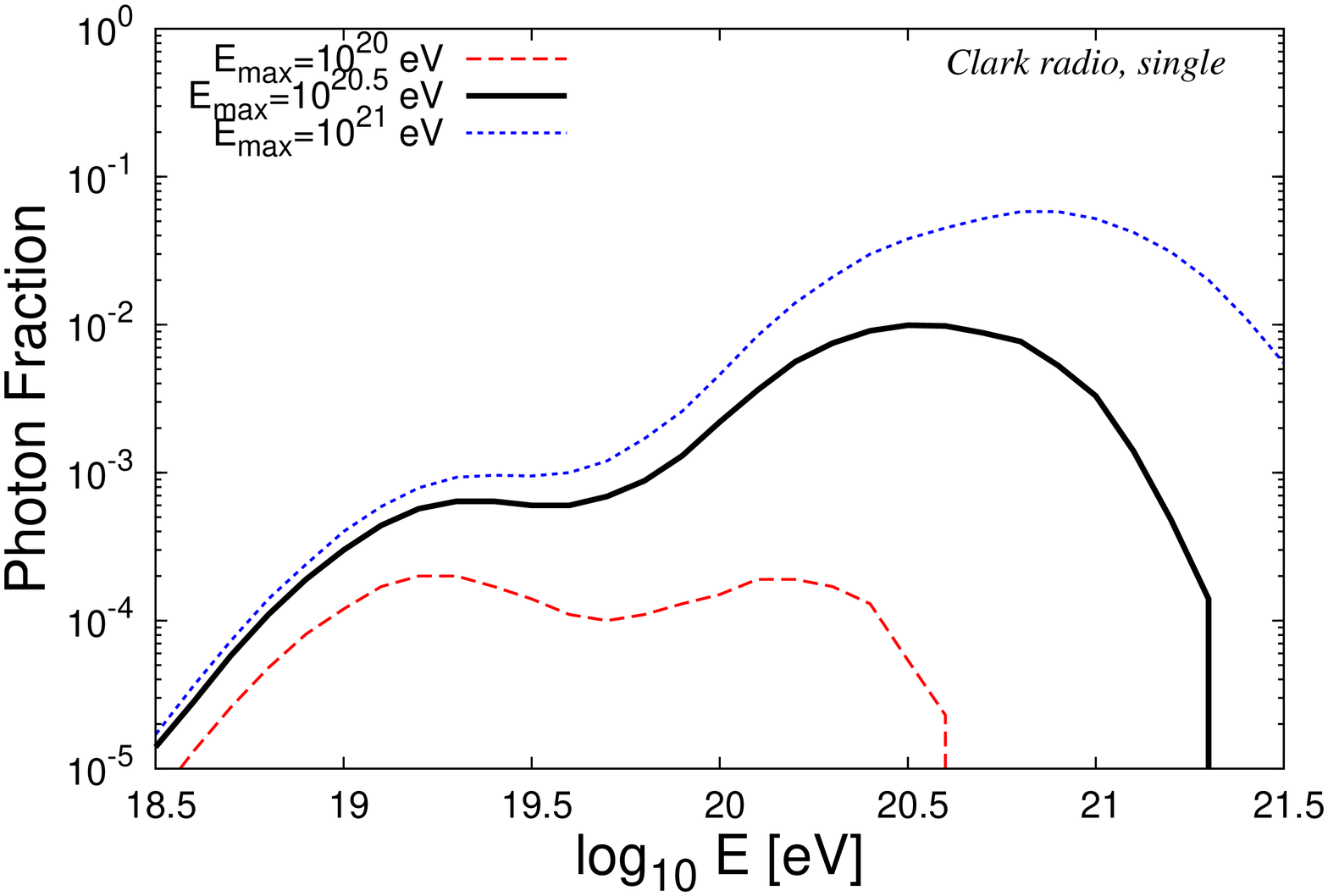}}}
\caption{{\bf Left}: The UHECR proton flux arriving from a homogeneous distribution of sources for different values of $E_{\rm max}$. {\bf Right}: The corresponding photon fraction of the arriving UHECR obtained from a homogeneous distribution of sources for different values of $E_{\rm max}$. For this plot, the label ``{\it Clark radio, single}'' refers to the assumptions about the radio background and magnetic field strengths, for which we have assumed the Clark et al. \cite{Parker:1970} radio background and extra-galactic magnetic fields stronger than 0.1~nG. The injected proton spectrum assumed for these results had a spectral index of $\alpha$=2.}
\label{Photon_Fraction_Cut-Off}
\end{center}
\end{figure}

In order to disentangle the effect on the photon fraction caused by the maximum energy of the proton, 
$E_{\rm max}$, from the local spatial distribution of UHECR sources to be discussed in this paper, we here 
explicitly obtain the photon fraction for the different $E_{\rm max}$ values of 10$^{20}$~eV and 10$^{21}$~eV. 
The alterations to both the GZK feature and the photon fraction are shown in the left and right panels of
fig.~\ref{Photon_Fraction_Cut-Off}. For comparison we also show the results obtained in the previous sections,
with $E_{\rm max}$=10$^{20.5}$~eV, in the solid line curves. For all these calculations a homogeneous distribution 
of sources was assumed.
In the left-panel of fig.~\ref{Photon_Fraction_Cut-Off}, the effect of the cut-off energy on the arriving UHECR 
spectrum is shown. An understanding of this effect comes from a consideration of the 
breakdown of this flux from the different source shells shown in the left-panel of 
fig.~\ref{Ensemble_Results_Breakdown}, with the cut-off energy primarily 
effecting the contributions from the more local shells which dominate the UHECR flux at the cut-off energy. 
In the right-panel of fig.~\ref{Photon_Fraction_Cut-Off}, the effect of the cut-off energy on the photon fraction
of the UHECR flux is shown. Interestingly, the photon fraction at all energies is seen to increase when the 
cut-off energy is increased.
A qualitative understanding of this follows from the fact that (for a given injection spectrum) as the cut-off 
energy is increased, the UHECR energy injected into particles above threshold for $p\gamma$ interactions is 
increased, leading to an increased injection of UHE photons. 

However, a more complete understanding as 
to why the photon fraction increases significantly around the cut-off energy, when the cut-off energy is
increased, can be obtained through a 
consideration of the photon and proton fluxes at the cut-off energy.
As the cut-off energy is increased from 10$^{20}$~eV to 10$^{21}$~eV, the proton flux at this energy decreases by 
nearly an order of magnitude (see the left-panel of fig.~\ref{Photon_Fraction_Cut-Off}). 
Furthermore, the photon flux at the cut-off energy increases by more than an order
of magnitude, resulting in more than a two order of magnitude increase in the photon fraction at the cut-off 
energy.
This large increase in the photon flux at the cut-off energy has two components.
The first of these may be understood by considering the photon flux
produced by protons injected in the 3-9~Mpc and 9-27~Mpc shells, the dominant contributing shells 
to the arriving photon flux at the cut-off energy. For a cut-off of 10$^{20}$~eV, the majority of the 
photon flux at the cut-off energy arrives from sources out to $\sim$10~Mpc (ie. only the first 
Mpc or so of the 9-27~Mpc shell actually
contributes), after which attenuation of the flux through pair production interactions
kills contributions from sources further away. However, as the cut-off energy is increased to 10$^{21}$~eV, 
the contributing region to the photon flux at cut-off increases to roughly half the 9-27~Mpc shell (the inner 
10~Mpc of the shell), due to the increase of the pair creation interaction length with energy 
(see the right-panel of fig.~\ref{Energyloss_Times}).
Such a contributing region argument leads to a factor of 2 increase in the photon flux.
The second, by far the dominant effect causing the large increase in the photon energy flux at
the cut-off energy, is the fact that the number flux from a given source at the cut-off energy
is the same for the different $E_{\rm max}$. This follows from the fact that the photon flux at 
the cut-off energy arrives predominantly from the first pions produced by protons after being emitted 
from the source (ie. whose distribution follows from $n_{\rm max}=0$ in eqn~(\ref{pion_contribution})).
Hence, the energy flux at the cut-off energy, for an order a magnitude in $E_{\rm max}$, increases
by an order of magnitude.
The overall effect on the photon fraction is a general increase as the cut-off energy is increased, 
with the most significant increase occurring for the second peak of the photon fraction curve, 
which peaks at approximately the cut-off energy, $E_{\rm max}$.

\subsection{A Local Enhancement/Depression of Cosmic Ray Sources}
\label{local_distribution}

In the previous section we focussed on the expected results obtained from a homogeneous distribution
of sources. In this section we depart from the homogeneous assumption to investigate how the photon fraction
is affected by a local source distribution.
In order to account for the natural fluctuations that must exist at some level within the homogeneous distribution
locally due to the fact that sources are not distributed continuously in space but are finite in number,
occupying particular positions, we here investigate the effect that local source 
over-densities/under-densities can have on the arriving photon fraction.

\begin{figure}[t]
\begin{center}
\rotatebox{-90}{\resizebox{5.0cm}{!}{\includegraphics{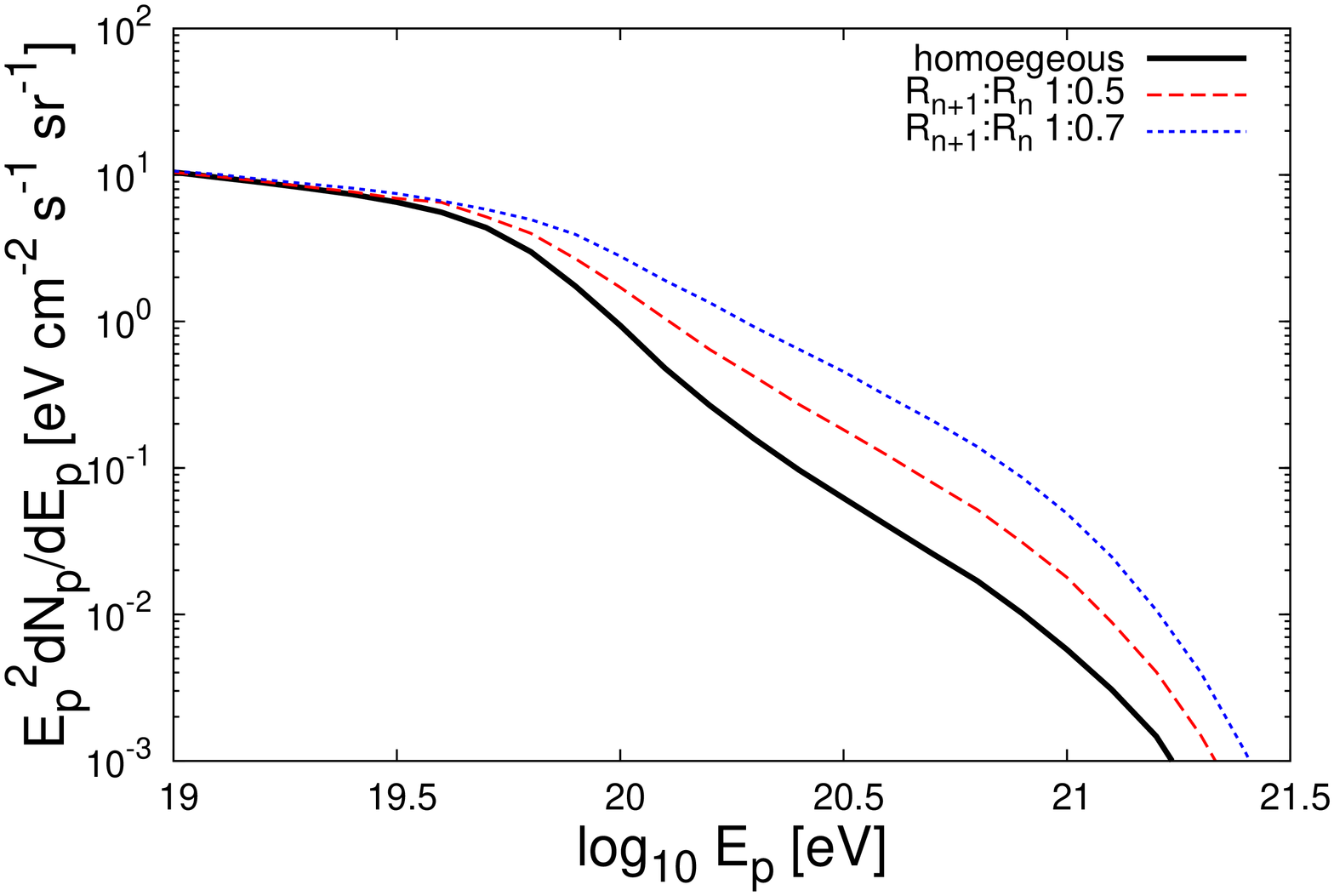}}}
\rotatebox{-90}{\resizebox{5.0cm}{!}{\includegraphics{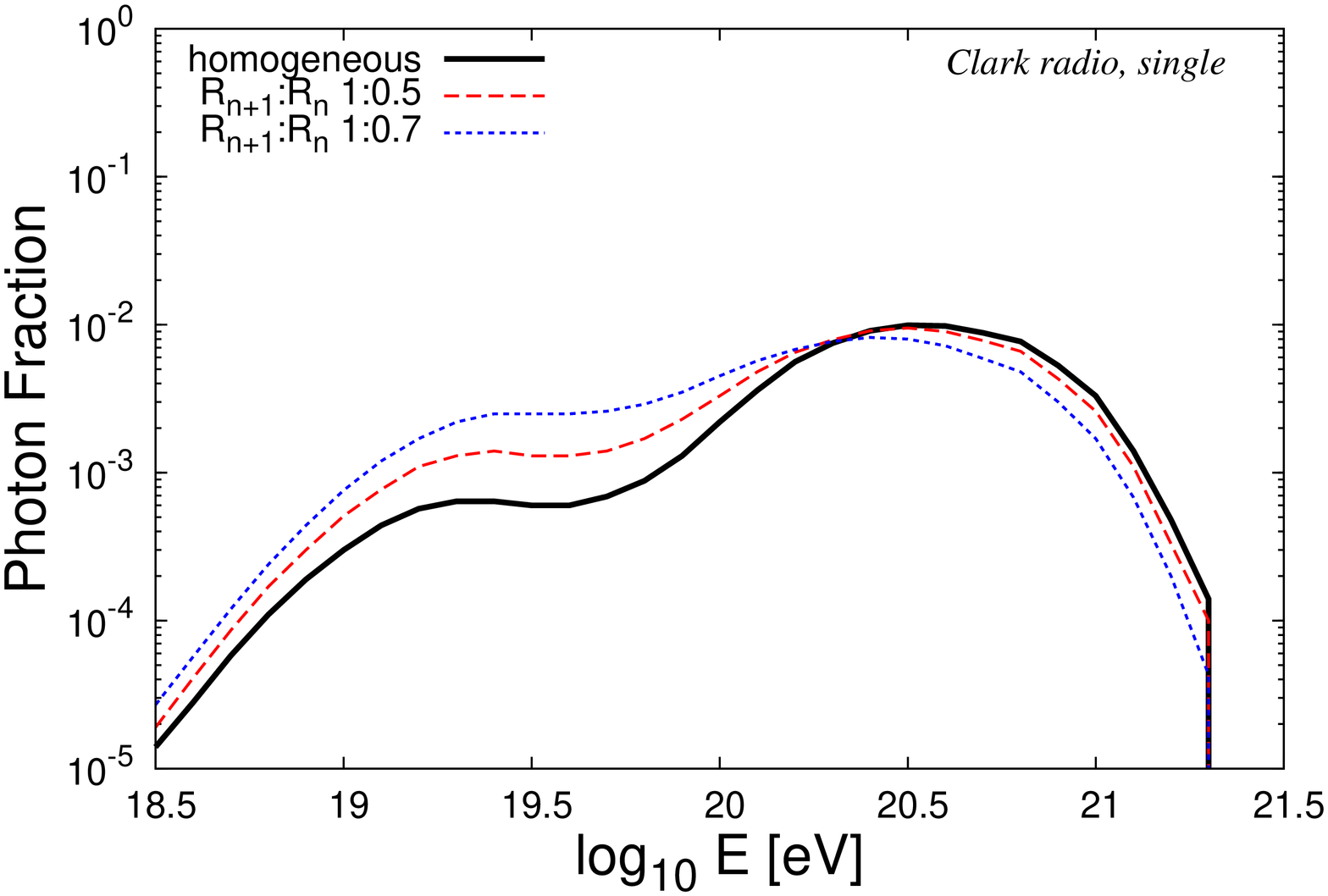}}}
\caption{{\bf Left}: The UHECR energy flux arriving from different distributions of sources for the case of a local source {\bf over-density}. {\bf Right}: The photon fraction of the UHECR flux from different distribution of sources for the case of a local source {\bf over-density}. For this plot, the label ``{\it Clark radio, single}'' refers to the assumptions about the radio background and magnetic field strengths, for which we have assumed the Clark et al. \cite{Parker:1970} radio background and extra-galactic magnetic fields stronger than 0.1~nG. The injected proton spectrum assumed for these results had a cut-off energy of $E_{\rm max}$=10$^{20.5}$~eV and an injection spectral index of $\alpha$=2.}
\label{Photon_Fractions_enhancement}
\end{center}
\end{figure}

In terms of over-densities, for the extreme case in which one of the shell regions dominates completely the 
arriving UHECR and photon flux above 10$^{19}$~eV, the proton and photon fluxes arriving to Earth would be those
from the corresponding shell in fig.~\ref{Ensemble_Results_Breakdown}. 
From these results, a large photon fraction is found to
be achievable for fluxes soley from either local or distant shells, though the energy at which this occurs
is very different in the two cases. This will be addressed in section~\ref{single_source}.
However, typically the expected contribution from each shell 
should scale approximately with the width of the shell, with contribution ratios of roughly 
1.0~:~0.3~:~0.09~:~0.03~:~0.01 being expected. 
If a local over-density of sources within the surrounding shells considered did exist, the arriving 
photon fraction and GZK feature would both differ significantly from the homogeneous source distribution 
scenario. Describing the ratio of the contributing shells as $R_{n+1}:R_{n}$ (with $n+1$ referring to further 
away shells), having a ratio of 1:0.3 for the homogeneous source scenario, we alter this ratio in order to 
simulate the effects of a local over-density, with nearby sources gaining a weighting factor when summing 
up the shell contributions.
In fig.~\ref{Photon_Fractions_enhancement}, both the effects this has on altering GZK feature 
(in the left-panel) and the alterations to photon fraction (in the right-panel) are shown. The GZK feature 
for this scenario is found to be somewhat delayed to higher energies, with a milder cut-off spectral slope, 
following the increase in dominance of the local contributing shell spectra, whose contributions can be 
seen in the left-panel of fig.~\ref{Ensemble_Results_Breakdown}. The photon fraction, however, alters only 
significantly at the lower energy peak, which increases in height with the introduction of the local UHECR 
source density enhancement, by more than a factor of 2. Such an increase in the photon fraction at lower 
energies approximately scales with the enhancement ratio of the source distribution locally, as might be
expected.

\begin{figure}[t]
\begin{center}
\rotatebox{-90}{\resizebox{5.0cm}{!}{\includegraphics{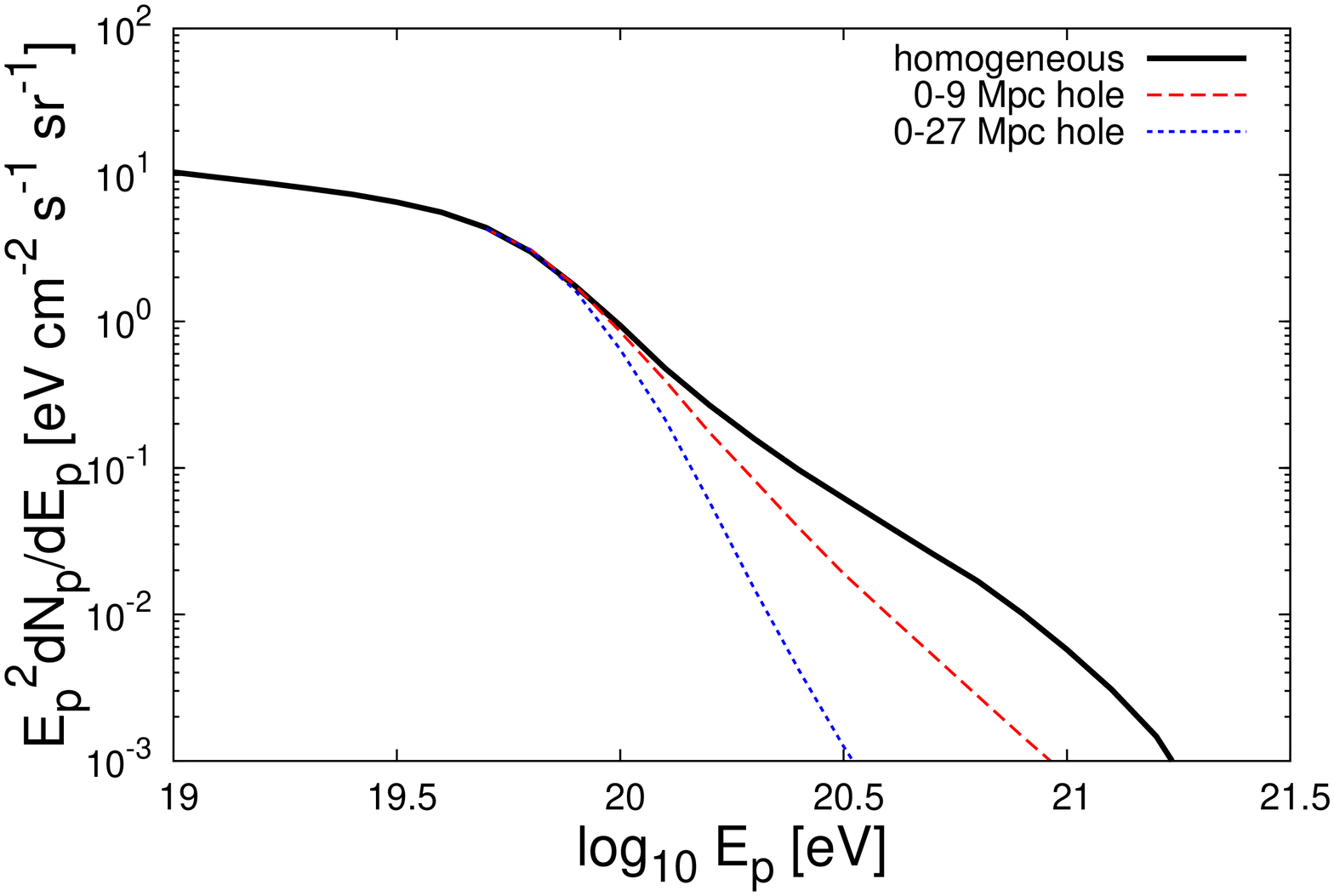}}}
\rotatebox{-90}{\resizebox{5.0cm}{!}{\includegraphics{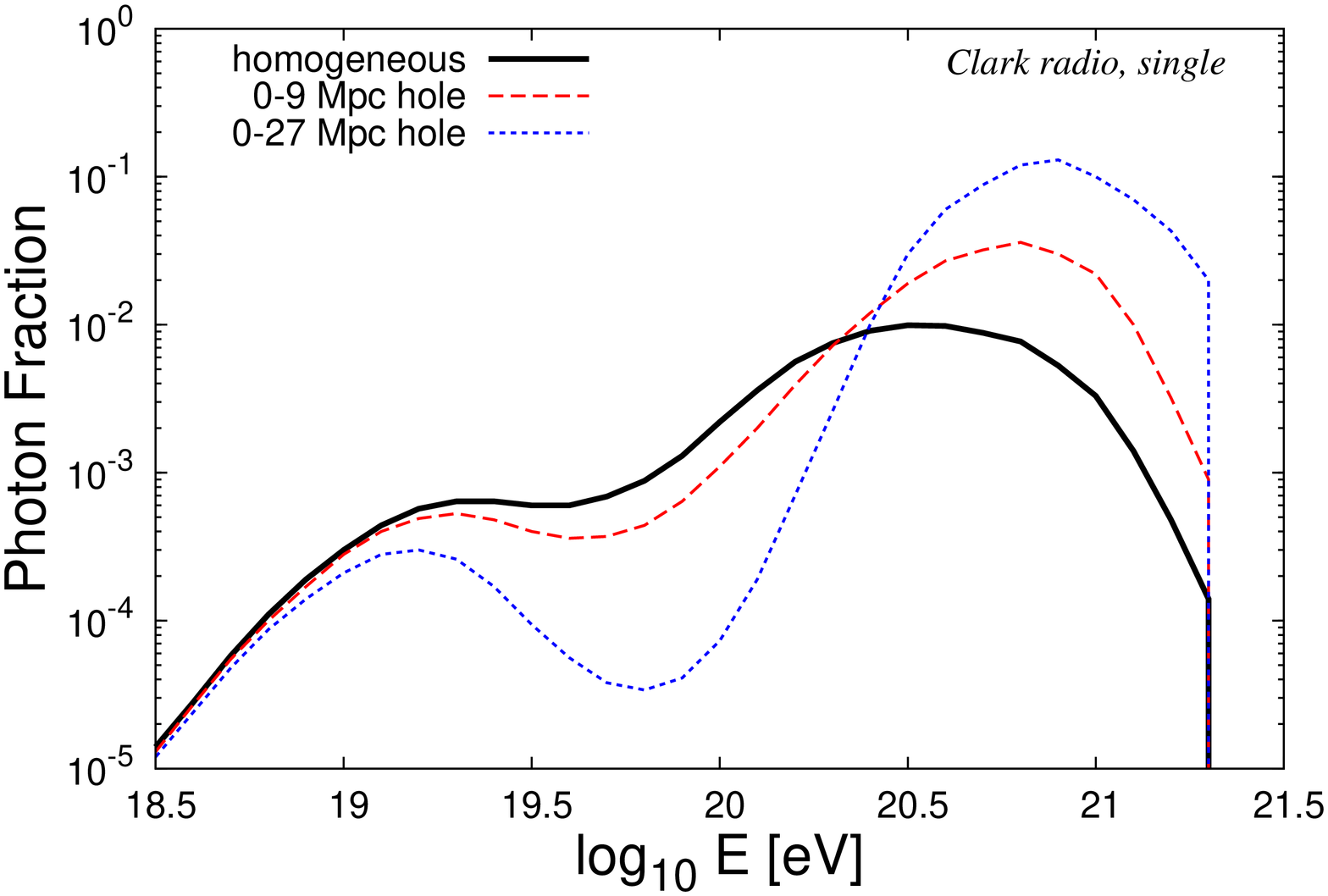}}}
\caption{{\bf Left}: The UHECR energy flux arriving from different distributions of sources for the case of a local source {\bf under-density}. {\bf Right}: The photon fraction of the UHECR flux from different distribution of sources for the case of a local source {\bf under-density}. For this plot, the label ``{\it Clark radio, single}'' refers to the assumptions about the radio background and magnetic field strengths, for which we have assumed the Clark et al. \cite{Parker:1970} radio background and extra-galactic magnetic fields stronger than 0.1~nG. The injected proton spectrum assumed for these results had a cut-off energy of $E_{\rm max}$=10$^{20.5}$~eV and an injection spectral index of $\alpha$=2.}
\label{Photon_Fractions_void}
\end{center}
\end{figure}

Following on in the vein that a departure from homogeneity might exist in the UHECR source distribution, we 
also attempt to account for the effect of a local depression in the source distribution by introducing a hole 
of UHECR sources surrounding Earth. The effects introduced to the GZK feature by a local depression of UHECR 
sources are shown in the left-panel of fig.~\ref{Photon_Fractions_void}, with the effects on the photon 
fraction being shown in the right-panel of fig.~\ref{Photon_Fractions_void}. Such a depression, in fact, 
is found to enhance the photon fraction of UHECR, with an increase in the hole size of the sources surrounding 
Earth, leading to more than an order of magnitude increase in the high energy end ($>$10$^{20.5}$~eV) of the 
UHE photon fraction plot. However, it should be noted that such a growth in the photon fraction only exists
in the region in which the energy flux of arriving UHE particles is less than 
4 orders of magnitude below that at 10$^{19}$~eV, so cannot be expected to be detectable by present UHECR
detectors. In this local depression of sources scenario, the GZK feature, contrary to the 
over-density of sources case, leads to a harder spectral cut-off feature. This is due to the dominance of 
the further away contributing shells, whose spectra are shown in the left-panel of 
fig.~\ref{Ensemble_Results_Breakdown}.

\subsection{The Photon Fraction from a Single Source}
\label{single_source}

\begin{figure}[t]
\begin{center}
\rotatebox{-90}{\resizebox{5.0cm}{!}{\includegraphics{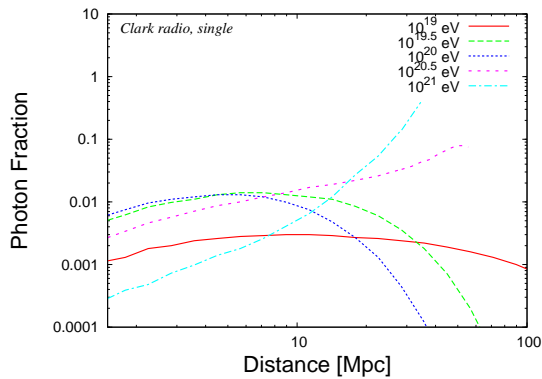}}}
\caption{The photon fraction of UHECR as a function of distance from the source for different arriving energies of photons/protons. For this plot, the label ``{\it Clark radio, single}'' refers to the assumptions about the radio background and magnetic field strengths, for which we have assumed the Clark et al. \cite{Parker:1970} radio background and extra-galactic magnetic fields stronger than 0.1~nG. The injected proton spectrum assumed for these results had a cut-off energy of $E_{\rm max}$=10$^{20.5}$~eV and an injection spectral index of $\alpha$=2.}
\label{photon_fraction_dist}
\end{center}
\end{figure}

Following on from these perturbations to the source distribution from the homogeneous scenario,
we here focus on the photon fraction of the propagating particles from a single UHECR source 
(assuming rectilinear propagation for protons). This provides an understanding of the photon
fraction expected in the extreme scenario of a single dominant local UHECR source. Though this 
scenario would seem to require uncomfortably slow diffusion in order to prevent violation of
anisotropy limits, it nevertheless can provide a useful insight to understand the results 
obtained from an ensemble of sources located at a particular radius.
We show in fig.~\ref{photon_fraction_dist} the evolution of the photon fraction as a function of
distance from the UHECR source, for different energy particles. The lines in the plot that terminate
early (ie. those for 10$^{20.5}$~eV and 10$^{21}$~eV particles) indicate that the energy flux from the source 
has decreased more than 4 orders of magnitude at the termination distance. Hence, though such a flux may contain
a large photon component, it would be unobservable to present generation UHECR detectors.
This photon fraction should be considered as a lower bound value, 
since the presence of magnetic fields will extend the actual propagated path
of the protons relative to that of the same energy photons (which themselves have a slight
path increase from the source due to their, much higher in energy, parent proton's deflected trajectory).

As a general rule, the photon fraction curves in fig.~\ref{photon_fraction_dist} for particles above the 
injection cut-off, $E_{\rm max}$, are different in shape than those below this energy.
For particles below $E_{\rm max}$, the photon fraction curve peaks at more local distances with
increasing energy particle. Furthermore, at energies below the peak value, the photon fraction 
is larger for higher energy particles. This follows simply from a consideration of both the
arriving proton and photon fluxes arriving from the different radii shells shown in 
fig.~\ref{Ensemble_Results_Breakdown}.
For particles above $E_{\rm max}$, the photon fraction in fact curves upwards (has a positive second
derivative) with increasing distance. Such an increase in the photon fraction follows from a consideration
of our analytic description to both the proton and photon fluxes. Above the energy, $E_{\rm max}$, at which the
injected protons are suppressed, the contribution to the proton fluxes, at a given energy, from
the pion losses of higher energy protons, are penalised. Since the photon fluxes of the same energy
originate from much higher energy protons, the suppression to the contribution of the photon flux from
multi-pion loss protons is also strong. Hence, at energies above the cut-off, the protons that arrive have
predominantly undergone no pion production processes and the photon flux arrives
predominantly from protons that have only undergone a single pion loss. From this, it follows
that the increase in growth of the photon fraction with distance for energies above $E_{\rm max}$
is the result of the decrease in the ability of the protons that have undergone no photo-pion
interactions to arrive at Earth and a lack of attenuation of the photons produced by protons that
have undergone only a single pion loss. Thus, this rise in the photon fraction, for particles
with energy greater than $E_{\rm max}$, only begins to terminate
once the photon flux becomes significantly absorbed through subsequent pair production interactions.


\section{The Effect of Cosmic Ray Diffusion on Results}
\label{diffusive_effects}

So far in the paper we have ignored the effect introduced by the presence of magnetic fields on the cosmic ray 
proton trajectories. Without such effects, both UHECR protons and photons propagate rectilinearly from the 
injection shell to Earth and hence the photon fraction arriving from each shell is as presented in 
fig.~\ref{Photon_Fractions} for the homogeneous distribution of sources assumption must be considered a 
lower bound. However, for consistency with the relatively strong extra-galactic magnetic fields of more than 
0.1~nG required for the termination of the $e/\gamma$ cascade, we here discuss the effects introduced by 
proton diffusion in the extra-galactic magnetic fields.

As discussed in ref.~\cite{Hooper:2006tn}, for Larmor radii significantly larger than the coherence scale of the
ordered magnetic field, a charged particle propagating through a uniform magnetic field
of coherence length, $L_{\rm coh}$, will undergo a deflection of size $\alpha$=$L_{\rm coh}/R_{L}$, where
$R_{L}$ is the Larmor radius of the proton. A particle traversing a distance $L$, through $L/L_{\rm coh}$
randomly orientated magnetic patches of size $L_{\rm coh}$, will therefore random walk in angle space, 
being deflected on average by
\begin{eqnarray}
\theta(E_{p})\approx\left(\frac{L}{L_{\rm coh}}\right)^{1/2}\alpha\approx 2.5^{\circ}\left(\frac{10^{19}~{\rm eV}}{E_{p}}\right)\left(\frac{L}{100~{\rm Mpc}}\right)^{1/2}\left(\frac{L_{\rm coh}}{1~{\rm Mpc}}\right)^{1/2}\left(\frac{B}{0.1~{\rm nG}}\right).
\end{eqnarray}
Such angular diffusion leads to the largest increase in path-length the proton propagates, compared 
to the field-free rectilinear distance, for the ``lower energy'' protons under consideration. The effect
on the proton propagation is to lead to an increased effective distance, $L_{\rm eff}$, which the protons 
propagate relative to the straight-line distance, $L$, with the relative increase given by,
$L_{\rm eff}/L \approx 1+\theta^{2}/2$. Thus, a nG strength extra-galactic magnetic field would
lead to a $\sim$10\% increase in a 10$^{19}$~eV proton's path-length for an $L$ of 100~Mpc.

For the homogeneous distribution of sources scenario, the contribution to the fluxes of protons and photons 
arriving at 10$^{19}$~eV from different source distances is shown in fig.~\ref{Ensemble_Results_Breakdown}. 
The arriving proton flux at 10$^{19}$~eV is seen to be dominated by 
sources beyond several hundred Mpc. Similarly the 10$^{19}$~eV photon flux, 
which results from roughly 10$^{20}$~eV proton losses, 
receives its dominant contribution from sources $\sim$100~Mpc at this energy. With
the photons propagating rectilinearly to Earth, and the effective proton propagation path being 
proportional to the inverse square of the proton's energy, the overall path extension of the two 
routes ($p$ propagation to Earth and $p\rightarrow \gamma$ and subsequent $\gamma$ propagation to 
Earth) is dominated by the proton's (ie. first route) path extension.
Thus, at 10$^{19}$~eV, the increase in path-length for the protons can be
expected to be significantly larger than that of the arriving photons. 
With the arriving flux from each radial source element
being constant for this scenario, the effect of diffusion is to weight the contribution from the more local 
sources relative to that from the more distant sources. Hence the effect of diffusion, for the homogeneous 
source scenario, is in fact similar to the case of a local source over-density. 
For a 0.1~nG field, a resulting mild, $\sim$10\%, decrease of the arriving proton
flux to the arriving photon flux at 10$^{19}$~eV, leads to a similarly mild increase in the photon
fraction $\sim$10$^{19}$~eV relative to the values obtained in the previous sections which 
ignored magnetic field deflection effects.


Should the UHECR protons be born into a region of large magnetic field, their Larmor radius will be
comparable or even smaller than the magnetic field coherence scale (which we here assume to be
1~Mpc). If this occurs, the cosmic rays propagation may become diffusive, and a consideration
of the magnetic field structure on smaller scales should be considered. If such a regime were
entered for UHECR close to the source region, the photon fractions expected could be
considerably larger than the values obtained here. In this respect, our photon fractions
results should be considered as lower bound values.

\section{Discussion}
\label{Discussion}

With increasing UHECR proton energy, the sudden onset of the pion-production process $\sim$10$^{19.6}$~eV
due to a combination of both the low energy resonance-like cross-section for this process and the sharply 
peaked nature of the CMB's 
black-body distribution leads to a very sudden decrease in the interaction length for UHECR protons propagating 
through intergalactic space and consequently a sudden growth of the secondary UHE photon flux produced 
through these interactions. This, in combination with the relatively large mass of the pion to that of the proton 
($m_{\pi}/m_{p}\approx$0.15), results in a sudden decrease also in the attenuation length of UHECR protons.
Despite the fact that the UHE photons produced also interact themselves with the CMB photons through a 
resonance-like cross-section, their interaction length in fact increases with energy due to the Klein-Nishina 
suppression of these interactions which occur with $s/(m_{e}c^{2})^{2}\sim 10^{3}$ (ie. $\gg 1$), where $s$ is the 
squared center-of-mass energy of the collision and $m_{e}$ is the rest-mass of the electron (the propagator in 
these interactions). As a consequence, the 
mean-free-path of UHE photons increases with energy, becoming comparable to the attenuation length of 
protons at 10$^{20}$-10$^{21}$~eV (depending on the extra-galactic radio background strength assumed), leading to 
the possibility of a relatively large photon content in the arriving UHECR flux at Earth.

The spectral shape of the GZK feature of the arriving proton flux has been demonstrated to have a very simple 
to understand origin through an analytic expression for the spatial and energy distributions of protons 
emitted by a single source, taking into account their subsequent cooling through photo-pion production 
interactions. Using the individual spectra from the different shells considered, the change in the spectral 
cut-off feature of the arriving UHECR spectrum was found for both a local over-density and under-density 
scenario. A local over-density of sources lead to a slight delay to higher energies of this cut-off feature, 
along with it having softer cut-off spectral shape. On the other hand, for the local under-density scenario, 
the introduction of a local hole of UHECR sources surrounding the Earth out to $\sim$10~Mpc, or even 
$\sim$30~Mpc, lead to a much steeper spectral cut-off feature being obtained.

The photon fraction results obtained, plotted as a function of UHECR energy, was shown to exhibit a double 
peaked feature. This feature was observed to be present in all 
scenarios considered 
(see figs.~\ref{Photon_Fractions},\ref{Photon_Fractions_enhancement},\ref{Photon_Fractions_void}) 
provided the extra-galactic magnetic field strength was sufficiently strong (more than $\sim$0.1~nG) that the 
$e/\gamma$ cascades were prevented from developing, with inverse Compton cooling being out-competed by 
synchrotron cooling for the UHE electrons. In all cases considered, the higher energy second peak peaked 
roughly at the cut-off energy of the proton injection spectrum, $E_{\rm max}$.
This being demonstrated to be the case explicitly in fig.~\ref{Photon_Fraction_Cut-Off} 
for several different values of maximum proton energy, $E_{\rm max}$.
For sufficiently large values of the maximum proton injection energy by the sources ($E_{\rm max}>$10$^{20}$~eV),
the higher energy second peak in the photon fraction was larger than the lower energy first peak, with
larger values of $E_{\rm max}$ increasing the height of this peak.
Furthermore, the introduction 
of a hole region around the Earth, void of UHECR sources, lead to a further increase in the photon fraction at this
peak value, at energies close to $E_{\rm max}$.
However, the introduction of a local over-density of sources into the nearby
(surrounding $\sim$250~Mpc) region was found to only mildly reduce this high energy peak in the photon fraction, 
increasing instead the lower energy peak.

With the prospect of even higher statistic rates at energies above the GZK cut-off with an Auger North detector,
information about the UHECR sources through the photon fraction information holds great promise. Already,
with the current Auger statistics, the photon fraction limits effectively exclude any, either homogeneous or 
local, topological origin scenario for UHECR \cite{Aharonian:1992qf,Berezinsky:1998ft}. 
Following these results, the next step will be the more difficult task of using the photon fraction 
information to probe the local UHECR source distribution. In this paper, we have investigated
possible results expected in this regard, demonstrating that photon fractions in the range 10$^{-2}$-10$^{-3}$
present realistic lower bound values.

\section{Conclusion}
\label{Conclusion}

The photon component of the UHECR spectrum has here been demonstrated to be a useful diagnostic
tool for:\\ 
\newline
{\em (1) The origin of a cut-off feature in the observed UHECR spectrum.}\\
Confirmation that the cut-off feature at the high energy end of the UHECR spectrum originates from
proton photo-pion interactions with background photons was shown to be promising for a range
of radio backgrounds, extra-galactic magnetic field strengths, and cut-off energies, as demonstrated in 
figs.~\ref{Ensemble_Results_Breakdown2},\ref{Photon_Fraction_Cut-Off}.\\
\newline
{\em (2) The local distribution of UHECR sources.}\\
In combination with the GZK cut-off feature, the photon fraction was shown to be a useful tool in the 
determination of the local UHECR source distribution through the consideration of local
source under and over densities, as demonstrated in 
figs.~\ref{Photon_Fractions_enhancement},\ref{Photon_Fractions_void}.

\acknowledgements{The authors would like to thank Paolo Coppi for his useful discussions on $e/\gamma$ cascades. AT acknowledges a research stipendium.}

\end{document}